\documentclass[final,leqno,onefignum,onetabnum]{siamltex1213}
\usepackage{graphicx}
\usepackage{amsmath,amssymb,amsbsy}
\usepackage{subfigure}

\title{Stability and breakup of liquid threads and annular layers in a corrugated tube} 

\author{Qiming Wang%Questions, comments, or corrections to this document may be directed to that email address.
}

\begin{document}
\maketitle
\slugger{mms}{xxxx}{xx}{x}{x--x}%slugger should be set to mms, siap, sicomp, sicon, sidma, sima, simax, sinum, siopt, sisc, or sirev

\begin{abstract}
%Documentation is given for use of the SIAM \LaTeX\ macros. These macros are now compatible with \LaTeX$2_{\varepsilon}$. Instructions and suggestions for compliance with SIAM style standards are also included. Familiarity with standard \LaTeX\ commands is assumed.
We study the stability and breakup behavior of an axisymmetric liquid thread which is surrounded by another viscous fluid layer through a long wave approximation. 
The two fluids are immiscible and confined in a concentrically placed cylindrical tube. The effect of the tube wall corrugation is taken into account in the model, which allows the access of the interaction between the wall shape and the thread interfacial dynamics. 
The linearized system is studied by the Floquet theory due to the presence of non-constant coefficients in the equation and the spectrum is computed numerically via the Fourier-Floquet-Hill method. The resulting features agree qualitatively with those obtained based on a lubrication model in the thin annulus limit, where the short-wave disturbances that would be stabilizing owing to the capillarity in the absence of wall corrugation, can excite some unstable long waves. Those results from the linear theory are also confirmed numerically by the direct numerical simulation on the evolution equations. Meanwhile, a transition on the dominant modes from the one for a tube without corrugation to the one with wall shape included is found. Finally the drop formation is shown in the nonlinear regime when the pinch off of the core thread is obtained. The annular film drainage regime is also approached slowly along with pinching when the tube wall is close to the thread interface.  
In addition, our results demonstrate the possibility to suppress pinching, depending on the averaged annular layer thickness and the variation in tube radius.
\end{abstract}


\begin{AMS}\end{AMS}

\pagestyle{myheadings}
\thispagestyle{plain}
%\markboth{TEX PRODUCTION}{USING SIAM'S \LaTeX\ MACROS}

\section{Introduction}\label{sec:intro}

Capillary instability of a liquid thread or jet has received continuous attention for decades since the classic work of \cite{Rayleigh1878}. 
For a freely suspended liquid thread immersed in another immiscible fluid, the work on linear stability analysis reveals that the system is linearly unstable under a relative long-wave length perturbation regardless of the viscosity difference \cite{Tomotika35, Chandrasekhar1961}, which only reduces the growth rate. To probe the nonlinear dynamics of a single liquid thread or jet, large amount of work in literature has been focused on the reduced one-dimensional model, considering the relative simplicity while the essential physics is retained. Following the early work of \cite{Lee} and \cite{Bogy}, \cite{Eggers93, EgDu1994, Eggers97} studied the drop formation and the pinching dynamics for a liquid jet by including both inertia and viscous effects, while \cite{DTP1995b} investigated the breakup dynamics for a Stokes thread by dropping the inertia. Both models give self-similarity solution on the breakup but with different scalings and scaling functions, which agree well with the experiments \cite{Cohen1999, MT2000, RRR2003}; See also other related experiments on thread dynamics and drop formations \cite{Stone1992jfm, SBN94, Kowalewski96}. 
For axisymmetric inviscid drops or jets, the local self similar solutions can not be captured by the reduced model due to the interface overturn.
Those local solutions from self similar dynamics are further confirmed numerically on the full axisymmetric problems; see \cite{ChenSteen1997, Day1998, LeppinenLister2003} for the inviscid jets or drops, \cite{WPB1999, AWB2002, ChenNotzBasaran2002} for the Navier Stokes jet simulations and \cite{Pozrikidis99} for the Stokes jet via boundary integral simulation. 
In particular, \cite{AWB2002} compared the results between the reduced one-dimensional model and the full axisymmetric simulation for various parameters and showed good agreement for certain parameter space, especially when the capillarity dominates.
Furthermore, those different regimes for the scalings on pinching are found to yield to the so-called Stokes flow regime in \cite{ListerStone98} where the new self-similar scalings appear, as the external fluids become important as pinching is approached, which makes it essentially a two phase flow problem. In addition, the inertia is shown to be less important compared to the viscous effect and capillarity near breakup, while the scaling for the axial velocity is obtained by including a non-local contribution based on the boundary integral calculation. 
%However, when the viscosity contrast is relatively large in the two phase problem, whether the self-similar solutions exist or not is subtle (see the numerical study in \cite{SierouLister2003}, \cite{SuryoDoshiBasaran2004} and related studies in \cite{BPSW2013,Wang2013}). 
A comprehensive review on liquid jet problem is given in \cite{Eggers97} and recently in \cite{Eggers08}.

Related to the liquid jet or thread problem, a core-annular flow arrangement like system has attracted considerable attention because of the potential wide applications in engineering and industry processes, such as the lubricated piping, emulsification technique, drug delivery in biological system. To develop a understanding on the behavior of the coating layer and core thread, the simplified system and asymptotic analysis are desired and have been employed widely by many authors. In the presence of a base flow, \cite{Hickox71} showed, based on the regular perturbation techniques for long wave length disturbances, that both capillarity and viscosity stratification are destabilizing for long waves for either axisymmetric and non-axisymmetric perturbation, with the former dominating. For the case of a less viscous annular layer ($\lambda=\mu_2/\mu_1<1$ with $\mu_1$ and $\mu_2$ the viscosity of the core and annular fluids respectively),  it is shown that the viscous stratification is stabilizing the destabilization of capillarity for a band of Reynolds numbers \cite{Hu89,Hu90,Chen90,Chen91,PCJ89}. Such stable band or window may disappear when the annular film is relatively thick or its viscosity is enhanced compared to the core fluid. When the film is sufficiently thin, Georgiou \emph{et al.} \cite{Georgiou92} studied the linear stability of the system analytically and extended the results in \cite{Hickox71} to intermediate wavenumbers, where it is found that the viscosity stratification is destabilizing (stabilizing) for $\lambda>1$ ($\lambda<1$).
The interest of study also has been extended to the weakly nonlinear regime in 
\cite{Frenkel87,Papageorgiou1990} and strongly nonlinear regime in \cite{KC94, Kerchman95} respectively, which track the evolution of fluid interface and predict many interesting features of the system, such as traveling wave and chaotic solutions. In the weakly nonlinear regime, the film evolution is governed by the Kuramoto-Sivashinsky (KS) equation where the linearly unstable disturbances are saturated by the flow (the fourth order spatial derivative), while in the strongly nonlinear regime, the additional capillary nonlinear term competes with the KS saturation mechanism and may amplify the linearly unstable waves. One related problem for a thin liquid layer flowing exterior to a cylindrical fiber was studied in \cite{CM2006} which includes the equation in \cite{KC94, Kerchman95} as the limiting cases when the ratio of layer thickness to the fiber radius is sufficiently small.
Qualitative agreement between those model results and the experimental work \cite{AO90,CM2006} is obtained. Related review on the core annular flows can be found in \cite{Joseph97}.

In the absence of the base flow, Goren \cite{Goren1962} performed the linear stability analysis on the annular layer coating the inner or outer cylindrical tube, where the long-wave length approximation was found to be unstable and his results recovered those in \cite{Tomotika35} in proper limits. The nonlinear transient evolution of a stationary annular layer was studied in \cite{Goren1964}, whose results indicated the formation of periodic toroidal collars which agreed with his experimental results. Hammond \cite{Hammond1983} used the lubrication approximation to derive a thin film equation that is valid for either a annular film coating interior or exterior of a cylinder to the leading order. The equation is in the special case of the one used in \cite{KC94,Kerchman95} where the base flow is dropped. Again collar and lobe formation is suggested to exist when the film drainage occurs which is accompanied with an infinite time singularity. More details of the long time dynamics are studied in \cite{LRKCJ2006} which shows, in addition, for sufficiently long domain length, the collar may slide by eating the neighboring small lobes gradually. Those lubrication model has the advantage over the full problem simulations in that the computation time is significantly saved especially for those long time dynamics. As shown in \cite{Pozrikidis92jfm} who used the boundary integral method to simulate the dynamics of liquid threads and annular layers, the computation is time consuming when the film thickness is very small, which prevent one from finding the long time solution behavior, let alone the proper scalings there. Furthermore, the simulations in \cite{KT2001,KT2002} are claimed to take orders of $3\sim 4$ weeks for a typical set of simulation on the core annular flows for the full Navier-Stokes system. 
When the annular film is relatively thick, as expected from theoretical studies, the linearly unstable waves grow beyond the linear and weakly nonlinear regime, which, similar to the liquid jet or thread problem, eventually leads to pinch-off of the core thread; see the simulations by
\cite{Pozrikidis92jfm,Hagedorn2004} in the full axisymmetric problem, and recently by \cite{Wang2013} in a long wave approximation for a two phase flow problem inside a tube (A related work on long wave modeling of viscous compound liquid threads can be found in \cite{CMP2005}, where the outer tube is absent). The model in \cite{Wang2013} provides a simple way to couple the core and annular fluid in the strongly nonlinear regime, as compared to the literature where the annular thin film model is usually decoupled from the core dynamics \cite{Frenkel87, Hammond1983, KC94, Kerchman95} or coupled through a nonlocal term in the weakly nonlinear regime in \cite{Papageorgiou1990}, in particular, through an integral with the Bessel or Kummer functions as the kernels. 
A more recent work on active core-annulus coupling for a two-phase flow inside a narrow tube can be found in \cite{DR15}, who derived a lubrication model through the so called integral-boundary-layer method.
%\cite{Hammond1983}, \cite{CM2006}, \cite{LRKCJ2006}

The theoretical study when the tube wall shape variation is taken into accounted has received relatively less attention, although in reality, the wall of a tube or channel often has cross section variation and in many experiments related to microfluidics, the two phase flow is affected by the topographic effect (see the work \cite{Link2004,Utada2005, Chris2007,BTB2008,Humphry2009}). Dassor \emph{et al.}\cite{DDC84} studied a two phase flow system in a symmetrically sinusoidal channel in two dimensional space in the case of small Reynolds number and small wall shape variation relative to the core thickness, where a wavy fluid interface shape is found that is dependent on the amplitude and a phase shift relative to the wall. 
Ransokoff \emph{et al.} \cite{RGR87} investigated the foam formation within a channel with various shapes of cross section which varies slowly in the axial direction, where it showed there exists a critical capillary number, below which, the breakup time is inversely proportional to the capillary number.
Gauglitz and Radke \cite{GR90} analyzed a similar problem with a liquid film coating a corrugation wall while surrounded by the gas phase, where a thin film equation with full curvature retained is derived, for which the numerical simulation is carried out to capture the nonlinear evolution. 
%Also the agreement on film rupture time with the experiments were shown, which is similar to the conclusions in \cite{RGR87}.
%\cite{RGR87}
Recent computational studies on this subject can be found in \cite{KT2001jfm, MK2006, Olgac2006}, but without detailed theoretical linear analysis. 

To our knowledge, Wei and Rumschitzki \cite{WR2002a} first carried out the systematic linear analysis for the core-annular flows with an asymptotically small corrugated tube wall, which shows the effect of the coupling between the wavenumber of the initial surface perturbation and the wall harmonics, which, in certain cases, can excite unstable long wave length disturbance. Meanwhile, short wave modes that would have been stable in the case for a straight tube with no corrugation can be unstable, while stable window for short waves can still be obtained when the wall has sufficiently large wavenumber.
In addition, \cite{WR2002b} performed a numerical study on the weakly nonlinear model which is a KS like equation. It is shown when the interfacial tension is comparable to the shear, the regular traveling wave solution is favored over the chaotic solution due to the wall effect. When the interfacial tension is dominating, the growth of unstable modes will go beyond the weakly nonlinear regime, which requires other model to capture the dynamics. This is our aim in current study. 

%Despite the fact that little work has been done for two-phase flow with topographic effect in theory, many experiments have been carried out in microfluidic devices \cite{Link2004,Utada2005, Chris2007,BTB2008,Humphry2009} where the wall shape appears to be a controlling or at least a contributing factor. In addition, other applications can be found in human biological system \cite{JBG2011,LHFG2014} where the vessel wall is usually modeled as elastic soft tube whose shape is varied. This further motivates our modeling studies.

In this paper, we employ a long wave theory to derive a set of evolution equations, which take into account the interaction between the core and annular fluids, as well as the annular layer and corrugated tube walls. By assuming a much less viscous annular fluid than the core, the viscous effect from the annular layer enters the leading order equations, which is a modification to the classical single jet model \cite{EgDu1994, DTP1995b}.
The stability of the threads and layers is then considered in both the linear and nonlinear regimes.
The rest of the paper is organized as follows: the governing equation as well as the asymptotic reduction is shown in \S\ref{sec:equations}, where a thin film equation is derived to show consistency with the model in literature. Linear theory based on numerical work and the nonlinear evolution of the model equations are shown in \S \ref{sec:results}. Finally, the concluding remarks are given in \S \ref{sec:concl}.

\section{Mathematical equations}\label{sec:equations}
\subsection{Governing equations}
\begin{figure}
   \centering
   \includegraphics[width=3.5in,height=2.in]{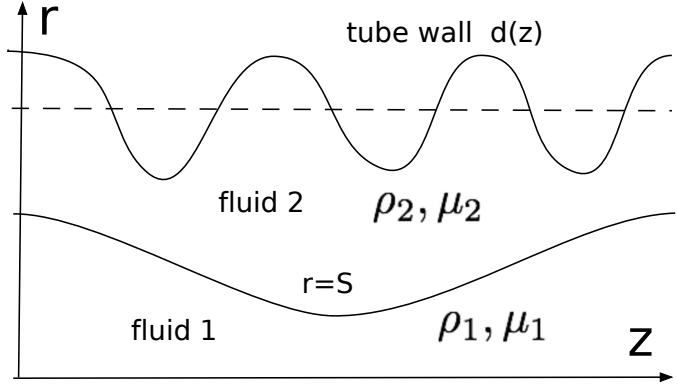} 
   \caption{The mathematical domain under consideration. The dashed line is the averaged tube radius. 
 Two viscous immiscible fluids, with densities $\rho_1,\rho_2$ and viscosities $\mu_1,\mu_2$ respectively, are separated by the interface $r=S$ and bounded by the tube wall $r=d(z)$.  For sinusoidal wall shape, the dimensionless tube radius is $d=d_0+\sigma \cos(k_wz)$ as we will discuss below, where $d_0$ is the position as illustrated as the dashed line, $\sigma$ is the amplitude of the waviness and $k_w$ is the wall wavenumber.}
   \label{domain}
\end{figure}

We consider the axisymmetric evolution of a viscous core liquid thread surrounded by another immiscible fluid inside a cylindrical tube of (dimensional) radius $b(z)$ that varies in the axial direction with corrugation. See figure \ref{domain}. The interface is given by $r=S(z,t)$ with constant surface tension $\gamma$ and the core region is occupied by fluid 1 with density $\rho_1$ and viscosity $\mu_1$, and the annular region is filled by fluid 2 with $\rho_2, \mu_2$. The flow fields are assumed to be axisymmetric with velocity components $(u,0,w)$ in terms of cylindrical coordinates $(r,\theta,z)$.

The dimensionless equations can be made if lengths are rescaled by the undisturbed core thread radius $a$, velocities by $\gamma/\mu_1$, time by $a\mu_1/\gamma$, pressure by $\gamma/a$. The subscripts $i=1, 2$ are hereinafter used to denote core and annular fluids respectively. The difference from the problem studied in \cite{Wang2013} is that the tube wall here has cross section variations, $d=d(z)=b(z)/a$ and for simplicity, the sinusoidal shape is considered in present work.
%The basic problem setting is similar to \cite{Wang2013} except here the tube wall is allowed to have topological structure, such as wavy wall or trench. To avoid difficulties, we simply assume $d=d(z)$ (with subscripts denoting partial derivatives) with $d_z<\infty$.  
Thus we have the governing equations
\begin{align}
&\chi_i R_e\left(u_{it}+u_iu_{ir}+w_i u_{iz}\right)=-p_{ir}+\lambda_i\left(\nabla^2 u_i-\frac{u_i}{r^2}\right),\label{momer}\\
&\chi_i R_e\left(w_{it}+u_iw_{ir}+w_i w_{iz}\right)=-p_{iz}+\lambda_i\nabla^2w_i,\label{momez}\\
& \frac{1}{r}\left(ru_i\right)_r+w_{iz}=0,\label{cont}
\end{align} 
where subscripts denoting derivatives, $\chi_1=1, \chi_2=\chi=\rho_2/\rho_1, R_e=\rho_1 \gamma a/\mu^2_1$ and $\lambda_1=1,  \lambda_2=\lambda=\mu_2/\mu_1$ , with boundary conditions at tube wall $r=d=b/a$ and at interface $r=S$ respectively
\begin{align}
&u_2(d,z,t)=w_2(d,z,t)=0,\\
&u_1(S,z,t)=u_2(S,z,t),\quad w_1(S,z,t)=w_2(S,z,t).
\end{align}
Notice that our definition of viscosity ratio is the same as in \cite{Hu90}, \cite{ListerStone98}, \cite{Pozrikidis99}, \cite{Wang2013}, where the small $\lambda$ corresponds to a case with a very viscous core and a relatively less viscous annulus, while in some other studies, such as \cite{SierouLister2003}, the viscosity ratio may mean the opposite case. To ease the discussion, we stick with our definition.

The stress balances at the fluid interface are given as
\begin{align}
&\left[\frac{\lambda_i}{1+S^2_z}\left(2S_z(u_{ir}-w_{iz})+(1-S_z^2)(u_{iz}+w_{ir})\right)\right]_1^2=0,\label{tsb}\\
&\left[-p_i + \frac{2\lambda_i}{1+S^2_z}\left(S^2_zw_{iz}-S_z(u_{iz}+w_{ir})+u_{ir}\right) \right]_1^2= \frac{1}{S\sqrt{1+S_z^2}}\left(1-\frac{SS_{zz}}{1+S_z^2}\right),\label{nsb}
\end{align}
where $[()]_1^2$ denotes the jump across the interface, that is, $()_2-()_1$. Finally, there is a kinematic boundary condition for the interface position $S$ that is given by
\begin{equation}
u_i(S,z,t)=S_t + w_i(S,z,t)S_z.\label{kin}
\end{equation}

% long wave model
\subsection{Long wave model}\label{lwmodel}
The aim here is to capture the physics with a set of simplified equations and we perform an asymptotic reduction of the above equations by assuming the axial length scale is much larger than the radial one with slenderness parameter $\epsilon=a/\mathcal{L} \ll 1$. Then we have the transformation $\partial_z\rightarrow \epsilon\partial_{z'}$ where $z'$ is the axial variable in long wave regime. In the rest of the article, the prime is dropped to simplify the notation. We shall also assume a weak topography with $\epsilon d_z\ll 1$ so that the dimensional slope of the wall is small.
Following \cite{DTP1995b} and \cite{Wang2013}, we introduce the following expansion 
\begin{align}
& u_i\sim u_i^0+\epsilon^2u_i^1+\cdots,\quad w_i\sim w_i^0/\epsilon+\epsilon w_i^1+\cdots,\quad p\sim p_i^0+\epsilon^2p_i^1+\cdots.
\end{align}

In the fluid region, in order to take both core and annular fluids into account, we take the annular fluid to be less viscous, $\lambda=\epsilon^2\lambda_0$ (see also the derivation in \cite{Wang2013}). This is different from the usual long wave model in \cite{EgDu1994}, \cite{DTP1995b} because here we also want to include the effect of tube wall by capturing the interaction between the tube wall and annular fluids. When there is no outer wall confinement, a similar limit $\lambda\ll 1$, where the core fluid thread is much more viscous than the surrounding medium, has been considered in the asymptotic modeling in \cite{ListerStone98,BPSW2013}.
We further ignore the inertial effect in the annular region, which amounts to assuming $\chi R_e<O(\epsilon^2)$.  This allows us to make analytic progress with the model.
From now on, we drop the superscript $0$ but retain $1$ to emphasize the high order term. The axial momentum equation then gives 
\begin{equation}
p_{2r}=0,\quad p_{2z}=\frac{\lambda_0}{r}\left(rw_{2r}\right)_r,  % _r is added
\end{equation}
whose solutions are readily to be obtained, by using continuity equation
\begin{align}
w_2&=\frac{1}{\lambda_0}\left(\frac{r^2}{4}p_{2z}+A\ln(r)+B\right),\\
u_2&=-\frac{1}{\lambda_0}\left(\frac{r^3}{16}p_{2zz}-\frac{r}{4}A_z+\frac{r\ln(r)}{2}A_z+\frac{r}{2}B_z\right)+\frac{C}{r},
\end{align}
where $A,\ B,\ C$ are functions of $z$ and $t$. 

In the core, $w_1=w_1(z,t)$ in the leading order, which is independent of radial coordinate. Using continuity equation, $u_1=-rw_{1z}/2$ as in the classical single fluid jet problem \cite{EgDu1994, DTP1995b}. Before considering stress balances at interface, we eliminate $B$ and $C$ by using no-slip and no-penetration boundary conditions at wall $d=d(z)$ to obtain
\begin{align}
w_2&=\frac{1}{\lambda_0}\left(\frac{r^2-d^2}{4}p_{2z}+A\ln(r/d)\right),\label{v20}\\
u_2&=-\frac{1}{\lambda_0}\left(\frac{(r^2-d^2)^2}{16r}p_{2zz}-\frac{r^2-d^2-2r^2\ln(r/d)}{4r}A_z\right)\nonumber\\
&+\frac{1}{\lambda_0}\left(\frac{d^2p_{2z}+2A}{4rd}(r^2-d^2)d_z\right),\label{u20}
\end{align}
where $d_z=0$ corresponds to the case in \cite{Wang2013}. Combining with the continuous velocity condition along $r=S$, $w_1=w_2$ and $u_1=u_2$, 
after some algebra we arrive at the following relation between $A$ and $p_2$,
\begin{align}
A& = -\frac{S^2+d^2}{4}p_{2z} + \frac{f(t)}{S^2-d^2},
\end{align}
where $f(t)$ is the quasi one dimensional force in the liquid thread (see also \cite{Renardy94, Renardy95, DTP1995a,DTP1995b}).
Therefore
\begin{align}
w_2=\frac{r^2-d^2-(S^2+d^2)\ln(r/d)}{4\lambda_0}p_{2z} + \frac{f(t)}{S^2-d^2}\ln(r/d).\label{axialv}
\end{align}

On the interface, the tangential and normal stress balances in leading order, read as
\begin{align}
&u_{1z}+w_{1r}^1+2S_z(u_{1r}-w_{1z})-\lambda_0 w_{2r}= 0,\label{tsblw}\\
&p_1-p_2+w_{1z}=\frac{1}{S}-\epsilon^2S_{zz}\label{nsblw}
\end{align}
where in normal stress balance, $\epsilon^2S_{zz}$ is retained as in \cite{EgDu1994}, to obtain the short wave cutoff in the linear stability region.
The higher order $w^1_{1r}$ is obtained from the axial momentum equation,
\begin{equation}
w^1_{1r}=\frac{r}{2}\left(\bar{R}_e(w_{1t}+w_1w_{1z})-w_{1zz}+p_{1z}\right),
\end{equation}
where $\bar{R}_e=R_e/\epsilon^2$ (so to ensure inertia does not enter the leading order in the annular region, $\chi \sim o(1)$ is needed; so we have assumed that the highly viscous fluid is also more dense than the less viscous one in our long wave theory). 
Substituting (\ref{axialv}) into (\ref{tsblw}) and using (\ref{nsblw}) leads to
\begin{align}
A = \frac{\bar{R}_e S^2}{2}\left(w_{1t}+w_1w_{1z}\right) -\frac{3}{2}\left(S^2w_{1z}\right)_z + \frac{S^2}{2}\kappa_z,
\end{align}
where $\kappa=1/S-\epsilon^2 S_{zz}$. With inviscid gas or vacuum instead of the viscous annular fluids, $A=0$ as in the single jet model \cite{EgDu1994}. 
%Meanwhile, we obtain the expression for the pressure gradient $p_{2z}$ that is used to determine the axial and radial velocities in (\ref{v20}) (or (\ref{axialv})) and (\ref{u20}),
%\begin{equation}
%p_{2z} = -\frac{2}{S^2+d^2}\left(\bar{R}_e S^2\left(w_{1t}+w_1w_{1z}\right) -3\left(S^2w_{1z}\right)_z + S^2\kappa_z\right).
%\end{equation}
Using the above results and kinematic condition (\ref{kin}) on interface, we reach the following equations for interface position $S$ and axial velocity $w_1$, namely,  
\begin{align}
&S_t+\frac{1}{2}Sw_z+wS_z=0,\label{1dS}\\
&\bar{R}_e\left(w_t+ww_z\right)=3\frac{(S^2w_z)_z}{S^2}-\kappa_z-2\frac{G(S,d)}{S^2} \left(\lambda_0w - \frac{f(t)}{S^2+d^2}\right),\label{1dW}
\end{align}
where the subscript $1$ in $w$ is dropped and
\begin{equation}
G(S,d)=\frac{S^2+d^2}{S^2-d^2-(S^2+d^2)\ln(S/d)}.\label{gfun}
\end{equation}
In general, $f(t)$ is unknown and determined by some global constraint. 

When $\lambda_0=0$ and $f(t)=0$, the model recovers the one in \cite{EgDu1994} for a single fluid jet.
In current study, we assume spatial periodic solution on $[-L,L]$, which gives
\begin{equation}
f(t)=\frac{\bar{R}_e\int_{-L}^LS^2\left(w_t+ww_z\right)dz+2\lambda_0\int_{-L}^LGwdz -\epsilon^2 \int_{-L}^LS^2S_{zzz} dz}{2\int_{-L}^L\frac{G}{S^2+d^2} dz}.\label{ft}
\end{equation}
To ease the discussion, for the rest of the paper, we further impose an even thread interface profile with respect to $z=0$ (also for $d(z)$), then $w$ is an odd function on $[-L,L]$ from (\ref{1dS}). This is consistent with the stability analysis in literature in the sense that the analysis starts with a sinusoidal perturbation on the thread interface.
Subsequently $f(t)\equiv 0$, which is used in the rest of study here and we leave $f(t)\ne 0$ case to future exploration.
%which is the case studied in \cite{Wang2013} for a straight tube with no corrugation. 
In addition, we notice that with $f(t)=0$ in (\ref{1dW}), the model equations (\ref{1dS}) and (\ref{1dW}) are formally the same as the equations in \cite{Wang2013}, even though the wall shape effect has been included in the present study. Meanwhile, neglecting the wall variation, the two phase slender jet model in \cite{ListerStone98} is recovered formally by taking $\bar{R}_e=0$ and rescaling $\lambda_0=\ln d\bar{\lambda}$ (namely, $\lambda=\epsilon^2\ln d \bar{\lambda}$) with $\bar{\lambda}$ order one quantity as $d\rightarrow \infty$. 
%This limit is missed in the model of \cite{CMP2005} which considered the rigid wall as a free surface as well, i.e. a liquid compound jet.
%But we notice the axial and radial velocities in annulus are different because of the presence of wall shape variations.

\subsection{Thin annulus limit}\label{thinfilm} % thin film
Before examining the fully nonlinear model, we bring up one interesting limit that many authors have investigated, the thin annular film limit \cite{Hammond1983, Papageorgiou1990, WR2002a, WR2002b}. Following \cite{WR2002a}, we write
\begin{equation}
d=1+\delta(1 + \sigma'\phi), \quad S = 1+\delta - \delta h,\label{ans}
\end{equation}
where $\delta\ll 1$, $\phi(z)$ is the wall shape function, $h(z,t)$ stands for the annular film thickness and $\sigma'=0$ corresponds to a uncorrugated wall or a straight tube case \cite{Hammond1983, Papageorgiou1990}. At the fluid interface, substituting (\ref{ans}) into (\ref{gfun}), one arrives at
\begin{align}
G \sim \frac{3(h-3\sigma'\phi)}{\delta^3 h^4} + O(\delta^{-2},\delta^{-2}\sigma').
\end{align}
Rescaling time as $\tau = \delta^3t$, the leading order equation for $h$ is obtained as a modified Hammand equation
\begin{equation}
h_{\tau} = -\frac{1}{12\lambda_0}\left(h^3\frac{h_z+\epsilon^2h_{zzz}}{1-3\sigma'\phi/h}\right)_z,\label{heqn}
\end{equation}
where Hammond equation (see \cite{Hammond1983}) can be recovered by taking $\sigma'=0$, $\epsilon=1$ and adsorb the factors $12\lambda_0$ into time. 
%In this case, inertia term drops off as $w\sim O(\delta^4)$ from the kinematic condition. 
A further simplified linearized equation can be obtained by taking
 $h=1+\alpha\xi$ in (\ref{heqn}). After some algebra, the first few terms in equation are obtained as
\begin{equation}
\xi_{\tau} = - \frac{1}{12\lambda_0}\left[\xi_{zz} + \epsilon^2\xi_{zzzz}+\left(\left(3\sigma'\phi+3\alpha\xi\right)\left(\xi_{z} + \epsilon^2\xi_{zzz}\right)\right)_z\right].\label{xieqn}
\end{equation}
Upon rescaling in time variable, this equation is the same as the linear equation in \cite{WR2002a} in the strong surface tension limit if $\alpha\ll 1$ (see their (5.3)) and recovers the weakly nonlinear one in \cite{WR2002b} for $\alpha\sim O(1)$  (see their (4.2)). In the derivation in \cite{WR2002a,WR2002b}, the viscosity ratio is $\lambda\sim O(1)$ while in our long wave model, $\lambda\sim O(\epsilon^2)$. Thus it seems that the different viscosity contrast (at least for $\lambda\le O(1)$) only affects the time scale factor in the thin film equation.

We will not repeat the results in \cite{WR2002a} in detail, but merely use their results to validate our method in computing the spectrum later in the linear stability analysis. We only focus on the long wave model (\ref{1dS}), (\ref{1dW}) in the rest of the discussion.

\section{Results of the long wave model}\label{sec:results}
In this section, we first investigate the effect of wall on the linear stability of the long wave model, (\ref{1dS}) and (\ref{1dW}) by using the so called Floquet-Fourier-Hill (FFH) method (see \cite{Deconinck2006} for details). After reporting the results for linear theory, the nonlinear stability of the long wave model is studied through a direct numerical simulation.

%This numerical method is more straightforward than the methods used in \cite{WR2002a}. 
%We first reproduced the results in \cite{WR2002a} and then proceed to apply the method to our long wave model (\ref{1dS}), (\ref{1dW}).

\subsection{Linear stability analysis}\label{linr}
It is easy to see $\bar{S}=1$ and $\bar{w}=0$ still serve as the base state in our problem without any base flow or external force field. For simplicity and based on the observation in \cite{Wang2013} that the inertia contributes little to the dynamics, only the inertialess case is studied for the linear stability and we probe the effect of inertia later by direct numerical simulations.
We set  $S= 1+ S'$ and $w=0+w'$, where $|S'|\ll 1$, $|w'|\ll 1$. For brevity, the prime is suppressed in the following analysis. The linearized system is reduced as following in the Stokes limit ($\bar{R}_e=0$),
\begin{align}
w_z = \left[\frac{3w_{zz}+S_z+\epsilon^2 S_{zzz}}{2\lambda_0 G(1,d(z))}\right]_z =
\left[\frac{-6S_{zt}+S_z+\epsilon^2 S_{zzz}}{2\lambda_0 G(1,d(z))}\right]_z = -2 S_t\label{llw}
\end{align}

%When $\bar{R}_e\ne 0$, 
%\begin{align}
%2\bar{R}_e \int_{-L/2}^z S_{tt} + 4\lambda_0 G(1,d)\int_{-L/2}^z S_{t} - 6S_{tz} + S_z + \epsilon^2 S_{zzz}=0,
%\end{align}
%or 
%\begin{align}
%\left[\frac{2\bar{R}_eS_{tt}+4\lambda_0 G(1,d) S_t-6S_{zzt}+S_{zz}+\epsilon^2 S_{zzzz}}{2\lambda_0 G_z(1,d(z))}\right]_z = -2 S_t
%\end{align}
Meanwhile, we consider the wall's variation as a sinusoidal profile, $d=d_0 + \sigma\cos(k_w z)$ where $k_w$ is the wavenumber of the tube wall. Recall we have assumed a weak topography, $\epsilon d_z\ll 1$, which leads to the condition $\epsilon \sigma k_w\ll 1$, i.e. $k_w\ll \left(\epsilon\sigma\right)^{-1}$. So the dimensionless wall wavenumber need not to be small provided $\sigma$ is not large (see \cite{SKK2007} for similar argument in a related problem).
For a straight tube, $\sigma=0$, then $G(1,d)$ is a constant and the growth rate is obtained by taking Fourier transform on (\ref{llw}) directly as in \cite{Wang2013}. The dispersion relation is then given by
\begin{equation}
\omega = \frac{k^2(1-\epsilon^2k^2)}{4\lambda_0G(1,d)+6k^2}.\label{grsoln0}
\end{equation}
\begin{figure}
 \centerline{  \includegraphics[width=6.5in,height=5.in]{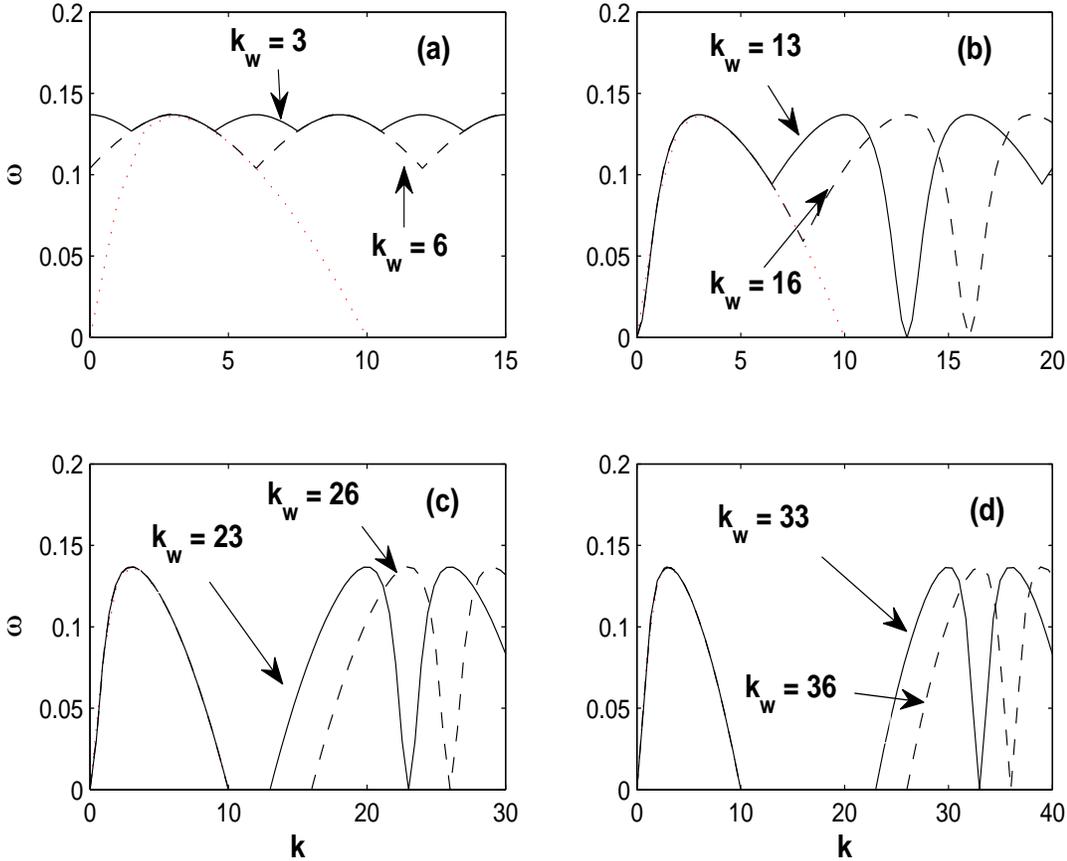} }
   \caption{For panels (a) to (d), the growth rate are plotted as a function of Bloch wavenumber $k$ with $d_0=5, \sigma=0.2, \epsilon=0.1, \lambda_0=1$ fixed. Solid lines are results for $k_w=3, 13, 23, 33$ while dashed lines for $k_w=6, 16, 26, 36$. The case with straight tube, i.e. $\sigma=0$ from (\ref{grsoln0}) is plotted together for comparison in dotted lines. For sufficiently large $k_w$, the growth rate recovers the straight tube case in the long wave regime $k<10$ in current case, or in general $\epsilon k<1$.}
   \label{linear}
\end{figure}
When the wall has cross section variation, $\sigma\ne 0$, the coefficient in (\ref{llw}) is a periodic function, which requires Floquet-Bloch theory to analyze the stability problem (see related studies in \cite{Papa87, WR2002a, BM04, BB06}).

We employ the FFH method \cite{Deconinck2006} for the eigenvalue problem here, which is a more straightforward numerical method than the ones in \cite{WR2002a}. We have first used this method to reproduce the results in \cite{WR2002a} and the convergence can be achieved quickly by using around $10\sim 20$ Fourier modes while \cite{WR2002a} reported that a relatively large linear system is needed to solve the eigenvalue problem accurately. The details of FFH for our problem can be found in the Appendix.

% derivation has gone to appendix
To calculate the spectrum numerically, we choose a cut-off $N$ on the number of Fourier modes of the eigenfunctions $\psi$ (see the Appendix), resulting in a linear system of dimension $2N+1$. As mentioned earlier, we reproduced the results in \cite{WR2002a} with around $N=5\sim 10$ (convergence within one period is achieved thus) and similar in the current linear problem. One advantage of FFH method is that one can handle any periodic function (in coefficient) in a general way by taking the Fourier transform (numerically), rather than simple cosine or sine function that can be easily incorporated in analysis. In current paper, we only focus on the simple sinusoidal profile though, which already gives a fairly complicated coefficient function $g$ in (\ref{eig1}).

%{\red Now we discuss the stability of the linearized problem.} 
We show typical results in figure \ref{linear}, where the wall shape wavenumber varies as indicated in figure. For comparison, the dotted lines in figure represent the growth curve in the case of uncorrugated tube, $\sigma=0$.
Unlike the presentation in \cite{WR2002a}, where the primary and secondary branches are identified, only the dominant growth rate or the largest eigenvalue is shown in figure.
Those results are in qualitatively good agreement with \cite{WR2002a}. When the wall shape has relatively long wave length variation i.e. small $k_w$, panel (a) shows that both the long and short waves are excited due to the presence of wall corrugation which leads to larger growth rate than the uncorrugated case for small and large $k$. For example, when $k=1, k_w=6$, the dominant growth rate will be associated with the first wall harmonic $|k-k_w|=5$, which grows faster than $k=1$ in the uncorrugated or straight tube case.
For intermediate range of $k$, the growth rate almost coincides with the uncorrugated case. As $k_w$ becomes larger, the overlap portion becomes larger, also for long waves, which is different from panel (a), the relatively small $k_w$ case. Furthermore, part of the unstable modes overlap the uncorrugated tube case completely as seen in the panel (c) and (d).
When the wall shape has sufficiently short wave modes, a stable band is observed for $1<\epsilon k<\epsilon k_w-1$ (see panel (c) and (d)), which appears periodically as well. One cutoff mode $\epsilon k_c=1$ is from the uncorrugated tube branch while the other is from the first wall harmonic $|\epsilon k_c-\epsilon k_w|$.
The result is similar to the findings in \cite{WR2002a}, which is as expected, since, although the lubrication model is only valid for asymptotically thin annular film, it in principle could reflect the basic features about the interaction between the capillarity and the wall harmonics.

The excitation mechanism persists even for sufficiently small $\sigma$ in our long wave model. Considering the small $\sigma$ limit, namely, $\sigma\rightarrow 0$ and setting $g=(4\lambda_0G)^{-1}$ with $\lambda_0=1$, an expansion of $g$ in $\sigma$ gives
\begin{align}
g&\sim \frac{1-d_0^2-(1+d_0^2\ln(1/d_0))}{4(1+d_0^2)} +\left(\frac{d_0^4+1-2d_0^2}{4d_0(1+d_0^2)^2}\right)\cos(k_w z)\sigma + O(\sigma^2),\nonumber\\
&\sim g_0 + g_1\sigma + O(\sigma^2)
\end{align}
Then the equation (\ref{eig1}) becomes 
\begin{align}
\omega\left(\tilde{S}-6g_0\tilde{S}_{zz}\right) + g_0\left(\tilde{S}_{zz}+\epsilon^2\tilde{S}_{zzzz}\right) - 6g_{1z}\omega\tilde{S}_z\sigma + \cdots =0
\end{align}
The first two terms of the above equation form the linearized equation for a straight tube ($\sigma=0$) for the dispersion $\omega$. Rewriting the above equation leads to
\begin{align} 
\omega \sim \frac{-g_0\left(\tilde{S}_{zz}+\epsilon^2\tilde{S}_{zzzz}\right)}{\left(\tilde{S}-6g_0\tilde{S}_{zz}\right)}\left(1+\frac{6 g_{1z}\tilde{S}_z}{\left(\tilde{S}-6g_0\tilde{S}_{zz}\right)}\sigma\right) + O(\sigma^2) +\cdots\label{grexpn}
\end{align}

Then it is clear that even very small corrugation amplitude would bring in a small correction term to the growth rate for straight tube case.  It is just that the time requires to see different behavior (due to corrugation) from the straight tube case is proportional to $\sigma^{-1}$ roughly (if the coefficient of $\sigma$ happens to be zero, then the correction comes from higher order terms in $\sigma$), namely, longer time is needed for small $\sigma$ to show the effect of corrugation. 
%(our simulation results also confirmed the transient behavior; see figure \ref{lineartest}). 

\begin{figure}
 \centerline{  \includegraphics[width=6.5in,height=2.5in]{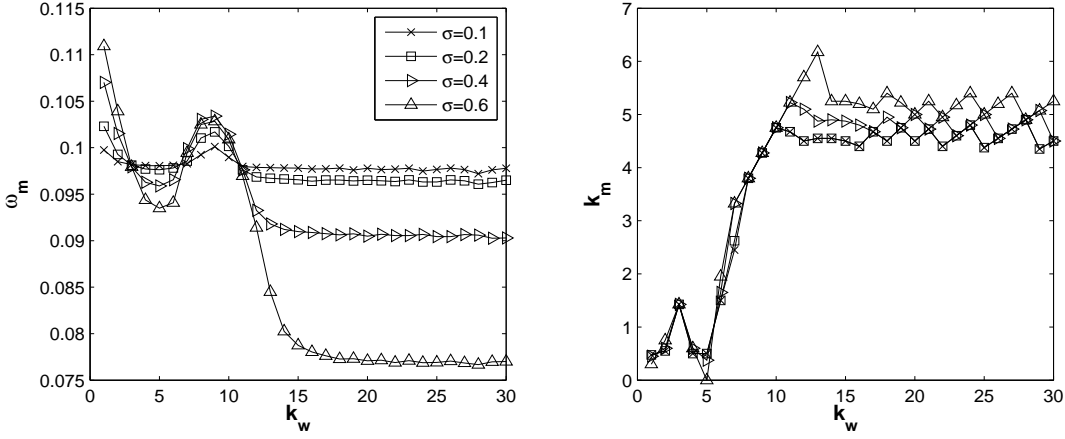} }
   \caption{The influence of wall wavenumber $k_w$ on the maximum growth rate $\omega_m$ and the associated wavenumber $k_m$ in the long wave regime for $d_0=2, \epsilon=0.1$.}
   \label{phs1}
\end{figure}
\begin{figure}
%   \centering
 \centerline{  \includegraphics[width=6.5in,height=3.in]{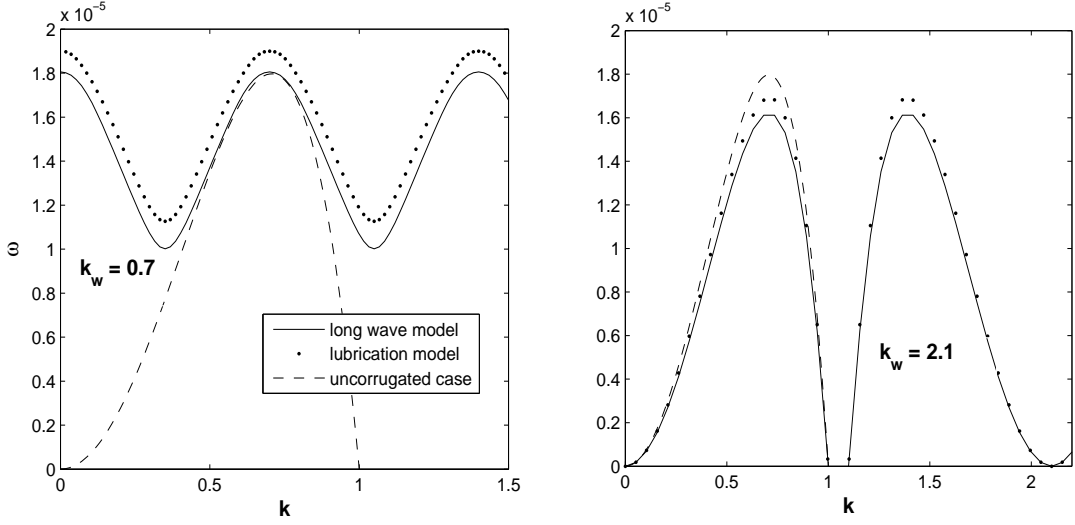} }
   \caption{Comparison between the results from linear theory based on (\ref{1dS}), (\ref{1dW}) in solid lines and on (\ref{xieqn}) in dotted lines. The dashed line illustrates the growth rate with uncorrugated wall. Here $d_0=1.1, \epsilon=1, \sigma=0.02 (\sigma'=0.2), \lambda_0=1$.}
   \label{linear_compr}
\end{figure}

The effect of wall wavenumber $k_w$ on the maximum growth rate $\omega_m$ is illustrated in the left panel of figure \ref{phs1} which shows that $\omega_m$ reaches a local minimum around $k_w\approx 5$ for $d_0=2$ that is close to the most unstable mode $k_m(\sigma=0,d_0=2)=k^0_m\approx 4.5$ in the straight tube case. This minimum value is also close to the uncorrugated case $\omega_m(\sigma=0)$ when $\sigma$ is small.
%Then it is clear that for relatively small wall wavenumber $k_w<10$ (or $\epsilon k_w<1$ in general), the maximum growth rate becomes larger than $\omega_m(\sigma=0)$ provided $k_w$ is deviated from $k^0_m$.
Meanwhile, the growth rate also tends to have a local maximum when $k_w\approx 0\ {\rm or}\ 9 (\approx 2k^0_m)$. 
%For $k_w\approx 9 (\approx 2k^0_m)$, $k_m\approx k^0_m$ as expected since $|k_m-k_w|\approx k^0_m$ should be the dominant mode.
 %For most cases, there are two locations where $\omega_m$ is reached within one $k_w$ period. We only highlight the one within the long wave range. The other one can be calculated easily by using $k_w-k_m$, since $|k_w-k_m-k_w|=k_m$ is the first wall harmonic of the initial perturbation. Thus, when $k_w$ is small, the mode other than the one in the right panel approaches $k^0_m$. For example, $k_m=0.5$ for $k_w=1$ from the right panel, then the other mode is 
% where $k_m\approx 0.5$ and $k_m\approx 4.5$ respectively, from the right panel of figure \ref{phs1}. Therefore, the wall harmonics are roughly $4.5$ which is close to $k^0_m$.
For a larger $\sigma$, this local minimum (maximum) has a smaller (larger) value. The parameter $d_0=2$ is chosen in figure \ref{phs1} but other choices of $d_0$ showed qualitatively similar results. 
%For example, the local minimum for small $k_w$ is seen around $k_w\approx k^0_m$.
For relatively large $k_w$, $\omega_m$ varies little and $\omega_m$ is seen to reach a smaller value for a larger $\sigma$. 
As will be shown in the nonlinear simulations later, the relatively large $k_w$ leads to a tube with an effective  radius $d_0-\sigma$. Therefore a larger $\sigma$ (with large $k_w$) means an effectively narrower tube which enhances the confinement and gives smaller growth rate. 

Meanwhile, we investigate the variation of the associated wavenumber $k_m$ at which $\omega_m$ is obtained in the right panel of figure \ref{phs1}. For most cases, there are two locations where $\omega_m$ is reached within one $k_w$ period. We only highlight the one within the long wave range or the smaller one. The other one can be calculated easily by using $k_w-k_m$, since $|k_w-k_m-k_w|=k_m$ is the first wall harmonic of the perturbation. 
For relatively small $k_w$, $k_m$ is found to be shifting between a small value close to zero and $k_w/2$ approximately.
%For relatively small $k_w$, $k_m$ is found to be located either around $k_w/2$ or close to zero. 
Corresponding to the local minimum from the left panel of figure \ref{phs1}, $k_m= 0.5$ is found in the right panel (for $\sigma=0.1,0.2$). Then the other mode $|k_m-k_w|=4.5$ is found to approach $k_m^0$. For relatively large $\sigma$ ($\sigma=0.4,0.6$ in figure \ref{phs1}), $k_m\approx 0$ so the other one is close to $5$ which indicates an obvious shift in the dominant modes.
%Corresponding to the local maximum at $k_w= 9 (\approx 2k^0_m)$, $k_m\approx k^0_m$ as expected since $|k_m-k_w|\approx k^0_m$ should be the dominant mode.
%For sufficiently small $k_w$, $k_m$ is close to $k_w/2$. For example, $k_m\approx 1.45, 1.55$ when $k_w=3$. 
%Meanwhile $k_m$ is seen to shift to a very small value at $k_w\approx 5$ which almost coincides with the most unstable mode $k^0_m$ in the straight tube case. 
%It is because this case leads to an almost zero wavenumber or the first wall harmonic. 
As $k_w>5$ increases, $k_m$ also increases until $k_w$ is sufficiently large, where $k_m$ oscillates around 5, which corresponds to the case when the unstable branch from the uncorrugated case has been recovered (see the similar case in the figure \ref{linear} for $d_0=5$). 
But there still exists small deviation in $\omega_m$ and $k_m$ from the exact value in the uncorrugated straight tube case.
%For $d_0=5$ in which case $k_m\approx 3$ for $\sigma=0$, we have observed $k_m$ is close to zero at $k_w\approx 3$ (data not shown). 
Similar deviation phenomenon in $\omega_m$ and $k_m$ was also pointed out in \cite{WR2002a}. 

%around the maximum value from the straight tube case, $0.0972$ in this case, $d_0=2$.

In figure \ref{linear_compr}, we compare the growth rates calculated directly from the long wave model (\ref{1dS}), (\ref{1dW}) with the results based on the lubrication model (\ref{xieqn}). In order to make the comparison, $\epsilon=1$ is taken and a thin annular layer is considered with $d_0=1.1$. Based on (\ref{ans}), $\delta=0.1$ and $\sigma=0.02\ (\sigma'=0.2)$ are used in the full long wave model. Suppose $\tilde{\omega}$ is the growth rate obtained from (\ref{xieqn}), the true growth rate in the long wave model is calculated as $\delta^3\tilde{\omega}$ which is plotted in figure \ref{linear_compr}. Two typical wall wavenumbers are chosen in figure \ref{linear_compr}, where the agreement is good, although the lubrication model seems to overestimate the growth rate slightly for this value of $d_0 (\delta)$ in the left panel, but in general has predicted the modes associated with the maximum growth rate.
%We also tested the case when the wall shape has multiple modes, where similar results are obtained as the single mode case reported here. However, we did not observe the asymmetric peak that is obtained in \cite{WR2002a}. 
In the next section, we carry out direct numerical simulations to further validate the results we obtained for the linear stability.

%The derivation is included in the Appendix.

\subsection{Nonlinear evolution} % nonlinear simulation
\begin{figure}
  \centerline{ \includegraphics[width=6.5in,height=4.5in]{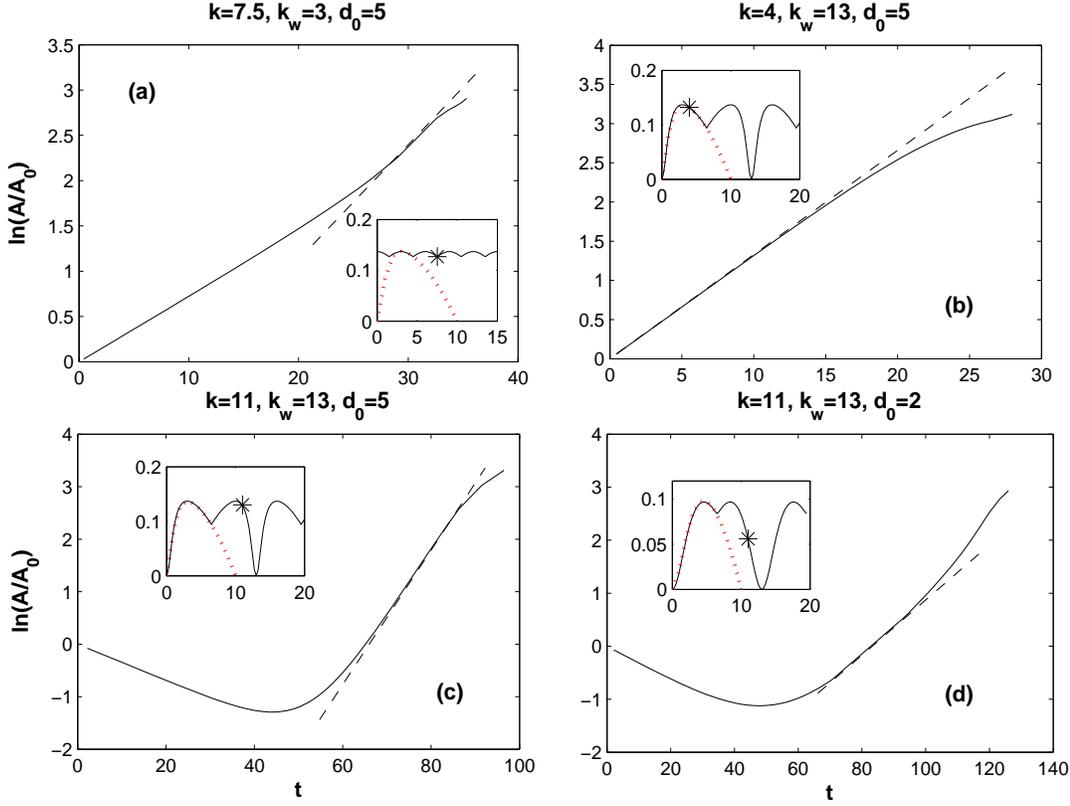} }
   \caption{Comparison between direct numerical simulation (in solid lines) and linear theory (in dashed lines) for various parameters as indicated in figure ($\epsilon k_w, \epsilon k$ would be the true wavenumber in the full axisymmetric problem). The quantity $A$ is defined as half of the difference between the maximum and minimum interface position. Meanwhile the Bloch wave number that is picked in simulations is highlighted in star symbol on the growth rate curves in the corresponding insets.
   Here $\epsilon=0.1, \lambda_0=1, \bar{R}_e=0, d_0=5, \sigma=0.2, A_0=0.05$ are used in simulations unless otherwise stated.}
   \label{lineartest}
\end{figure}
In the results reported here, we consider the wall shape as $d=d_0 + \sigma\cos(k_w z)$, with initial and (no axial flux) boundary conditions given by
\begin{align}
& S(z,0)=1 + A_0\cos(k z),\quad w(z,0) = 0,\\
& S_{z}(-L,t)=S_{zzz}(-L,t)=S_z(L,t)=S_{zzz}(L,t)=0\nonumber\\& w(-L,t)=w(L,t)=0.
\end{align}
%although we realized that some of the solution behaviors discussed here depend on the initial conditions as well.
To ensure both the wall and interface are periodic in the computational domain, the period length is taken as $L=\pi/\beta$ where $\beta$ is the common divisor such that $(k,k_w)=(n_k, n_w)\beta$ with both $n_{k,w}$ integers (see also \cite{WR2002b}). 
In addition, $\epsilon=0.1$, $\lambda_0=1$ and $A_0=0.05$ are fixed in simulation unless otherwise stated. Since $A_0=0.05$ is small, the early stage evolution is expected to follow the linear theory which will be confirmed numerically later.
The nonlinear equations are solved by the solver EPDCOL \cite{KM91} which has been successfully used for related problems (\cite{CMP2002} e.g.), and our previous studies (\cite{Wang2012, Wang2013}). This solver uses the finite element discretization in space and advances in time through the Gear's method. In the results reported here, about $1600\sim 3200$ space points are used within the computational domain. The simulation is stopped when the neck of the core thread $S_{min}<0.003$ or the vertical gap between the wall and interface is smaller than $0.0008$, which indicates a touching solution or annular film rupture rather than pinching of the core liquid thread in our problem.

\begin{figure}
  \centerline{ \includegraphics[width=6.5in,height=4.in]{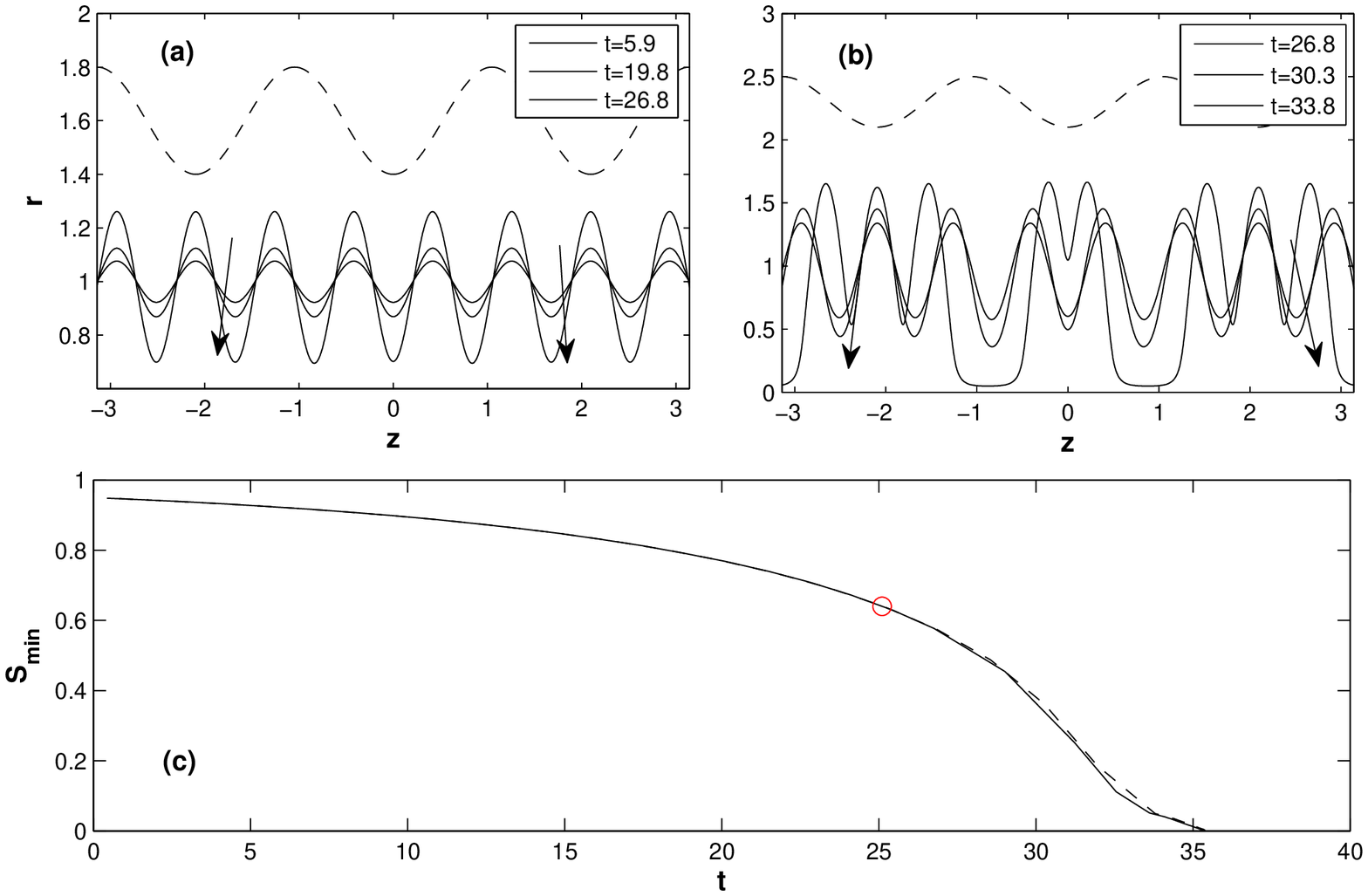} }
   \caption{The wall wavenumber is chosen the same as in panel (a) in figure \ref{lineartest}, namely, $k_w=3, k=7.5$. The other parameters are $\epsilon=0.1, \lambda_0=1, \bar{R}_e=0, d_0=5, \sigma=0.2$. In the upper panels (a) and (b), the dashed line only illustrates the wall shape but does not reflect the true position in simulation. It is seen that some longer waves dominate eventually over the initial short wave perturbation. In doing the simulation for this particular case, $L=2\pi$ is chosen to ensure the periodicity of wall and interface but only $[-\pi,\pi]$ is shown in figure. The circled point in bottom panel (c) indicates the time at approximately $t=25$, after which the evolution is dominated by another longer wave length perturbation, $|k-k_w|=4.5$. Dashed line in bottom panel is for the results with inertia included, $\bar{R}_e=1$ which only shows minor difference toward the end of evolution.}
   \label{evln1}
\end{figure}

\begin{figure}
  \centerline{ \includegraphics[width=6.5in,height=4.in]{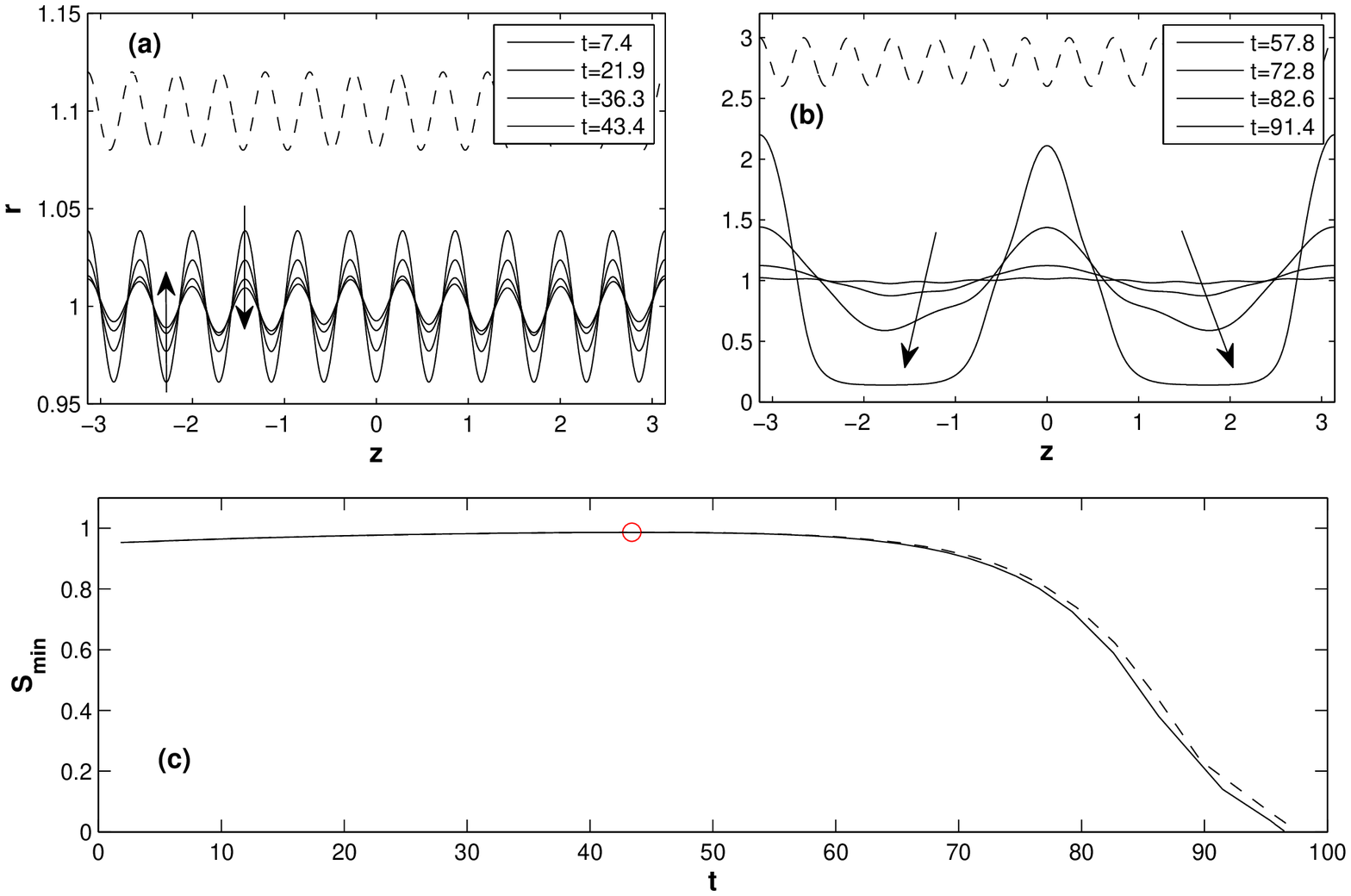} }
   \caption{The wall wavenumber is chosen the same as in panel (c) in figure \ref{lineartest}, namely, $k_w=13, k=11$. The other parameters are $\epsilon=0.1, \lambda_0=1, \bar{R}_e=0, d_0=5, \sigma=0.2$. In the upper panels (a) and (b), the dashed line only illustrates the wall shape but does not reflect the true position in simulation. It is seen that the relative short wave perturbation saturates at early stage but the perturbation eventually grows. The circled point in bottom panel (c) indicates $S_{min}$ reaches a maximum at approximately $t=43.4$, after which the evolution is dominated by another long wave length perturbation. Dashed line in bottom panel is for the results with inertia included, $\bar{R}_e=1$ which only shows minor difference toward the end of evolution.}
   \label{evln2}
\end{figure}

Our numerical results by choosing an initially slightly perturbed thread interface show fair agreement with the linear theory for which we solved by using the FFH method. A few typical results are shown in figure \ref{lineartest}. In the panel (a), the evolution is shown for $k=7.5, k_w=3, d_0=5$, in which case the growth rate is larger than the straight tube case based on the inset, where the star symbol indicates the parameter values used in simulation. 
In fact, due to corrugation, the disturbance is modulated by exp$(i(k\pm nk_w))$ ($n=1,2,\cdots$) and the eventual dominant growth rate is associated with the mode (first wall harmonic) $|k-k_w|=4.5$ (or $4.5\pm nk_w$ due to the periodicity). 
It is shown that the interface evolves first as if there were no wall roughness and the growth rate follows the value in the straight tube case, for $k=7.5, \sigma=0, d_0=5$ at early stage. After $t\approx 25$, the interface recognizes the wall topography effect and then follows the new dominant growth mode that is predicted by our FFH method. The agreement is shown by viewing the solid line from the direct simulation and the dashed line from linear theory agree with each other well, even though in this particular case, the time period showing the agreement is relatively short and the amplitude of the perturbation grows to finite for which the linear theory is not strictly valid. Our results do not indicate that the linear theory is able to predict finite amplitude growth. The agreement merely coincides with the fact that the linear theory usually works well even beyond its valid domain.
More details of the interface evolution will be revealed in figure \ref{evln1}. 
The important message that our numerical results delivered is the existence of the transition (see also the simulations in \cite{WR2002a}). It is as expected and can be explained using our (\ref{grexpn}), which indicates that the growth rate $\omega(\sigma=0)$ from the uncorrugated case dominates in short times and there exists a critical time when the correction term due to corrugation eventually becomes the same order as $\omega(\sigma=0)$.
%For the same $(k, k_w)$ but a smaller $d_0=2$ e.g., the transition occurs earlier at $t\approx 22$ and $\ln(A/A_0)\approx 1.6$, comparing to $t\approx 25, \ln(A/A_0)\approx 1.8$ for $d_0=5$ in panel (a), and hence, the period showing agreement could be longer (data not shown).
% delete
%\sout{This transition occurs for other Bloch wavenumbers as well, which is different from the results in \cite{WR2002a}, which seems to suggest that the interface follows the largest growth rate right from the beginning of the evolution if the perturbation is a relative long wave (see their figure 5(a)), even though the dominant mode comes from the branch rising from the interaction between the capillarity and wall harmonics. In our problem, it seems that there exists the transient growth as long as the growth rate differs from the value in the uncorrugated case.}
%delete
The transition exists for most of cases as long as the growth rate differs from $\omega(\sigma=0)$. But it will not appear if the thread experiences breakup before reaching the critical time.

In panel (b) and (c), we fix $k_w=13, d_0=5$ but choose two different wavenumbers for the initial condition: one corresponds to a mode that is roughly equal to the uncorrugated tube case, $k=4$ while the other one corresponds to a mode that would have been stable if the wall is uncorrugated, $k=11$. 
Then it is not surprising that the interface follows a single growth rate from the linear theory in panel (b). In panel (c), we again observe some transient growth first. At early stage, the relatively shortwave perturbation saturates (the growth rate is negative for $k=11$ in the straight tube case) and after $t\approx 43.4$, the evolution follows the growth rate from the linear theory for $\sigma\ne 0$. Panel (d) illustrated a similar case as panel (c) but with a smaller $d_0$, i.e. the wall is initially closer to the thread interface, which shows qualitatively the same behavior as panel (c). Such transient growth from saturated shortwave perturbations to unstable long wave ones is also observed in \cite{WR2002a}, in which the annular layer is asymptotically thin and the core dynamics is neglected.

We also carry out a close inspection in the case in panel (a) and (c) of figure \ref{lineartest} by showing the interface profiles in figure \ref{evln1} and \ref{evln2} respectively. Corresponding to the case in panel (a) of figure \ref{lineartest}, the interface evolves first as the case with no corrugation before $t\approx 25$ which is shown in the panel (a) of figure \ref{evln1}, while the panel (b) shows the evolution after the transition, which, despite the fact that some short waves surf on the large wave crest, follows longer wave length perturbation compared to the initial relatively short wave length perturbation. In particular, the eventual dominant mode is $|k-k_w|=4.5$ instead of $7.5$ as discussed above. The wall shape is illustrated in the upper panels only in order to view the relative phase between the interface shape and the wall. For the case in panel (c) of figure \ref{lineartest}, the transition from stabilizing to destabilizing is obvious. As indicated in figure, we plotted the thread profiles before and after the turning point $t\approx 43.4$. In the panel (a) of figure \ref{evln2}, the perturbation is indeed saturated as the amplitude of the perturbation decreases, while in the panel (b), the perturbation is amplified obviously and the core tends to pinch off eventually. This result is in good qualitative agreement with those in \cite{WR2002a}.
%The wall shape is illustrated in the upper panels only in order to view the relative phase between the interface shape and the wall.
The corresponding evolution of $S_{min}$ in bottom panel (c) further confirmed this transient evolution, where the circle indicates $S_{min}$ reached a maximum as time advances.

The dashed lines in the panels (c) of figure \ref{evln1} and \ref{evln2} are the results when inertia terms are included in calculations. Similar to the findings in \cite{Wang2013}, the inertia only affects the transient evolution slightly while having little effect both in the early stage and near pinching. It is also consistent with the scaling analysis in \cite{ListerStone98} where inertia is shown to give way to viscous fluid-fluid interaction when the region is slender.
Therefore, we focus mainly on the Stokes flow problem, i.e. $\bar{R}_e=0$, for the discussion here.

% drop formation
\subsection{Drop formation and film drainage}
\begin{figure}
  \centerline{ \includegraphics[width=6.in,height=4.in]{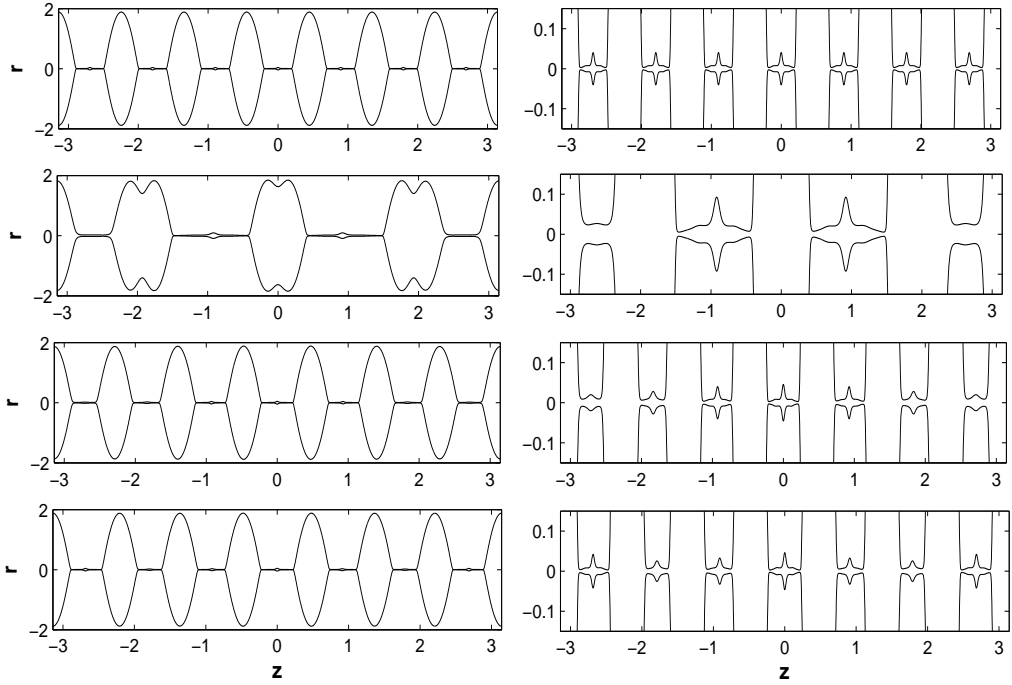} }
   \caption{Drop formation on the breakup when $d_0=5$ for different wall wavenumbers. Left column: The first panel shows the case with a straight tube $\sigma=0$. For the next three panels, $k_w=3, 13, 23$ respectively. The tube walls are out of sight in this figure but the shapes are similar to figure \ref{drop2}. 
Other parameters are $\epsilon=0.1, \sigma=0.2, \lambda_0=1, \bar{R}_e=0$. The panels in the right column show the corresponding satellite formations. The Bloch wavenumber in the initial condition is also fixed for all panels as $k=7$. The simulations are terminated at $t\approx 33.2, 32.4, 33.1,33.2$ for the four cases respectively, from top to bottom.}
   \label{drop1}
\end{figure}

\begin{figure}
  \centerline{ \includegraphics[width=6.in,height=4.5in]{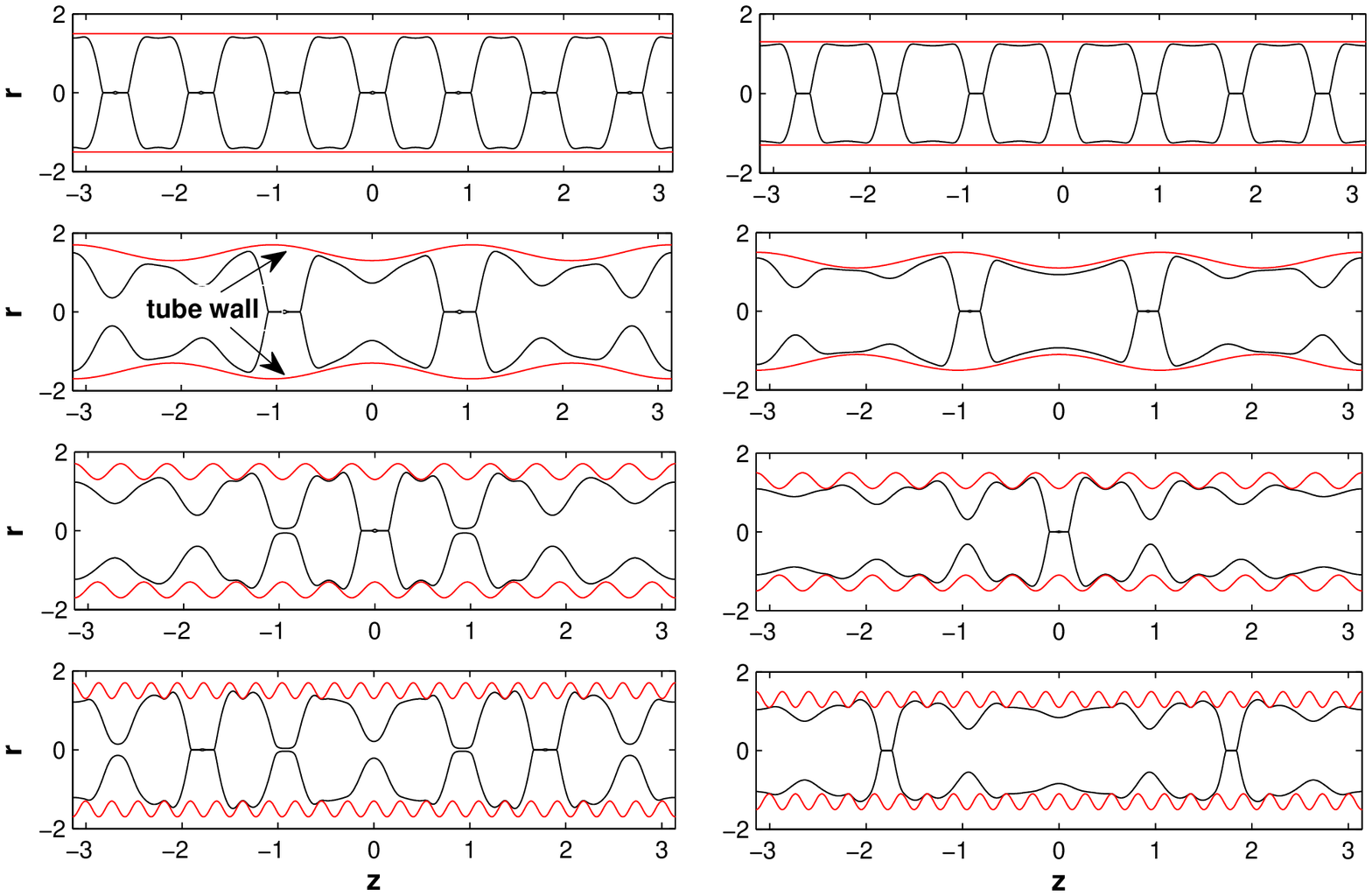} }
   \caption{Drop formation on the breakup when $d_0=1.5, 1.3$ in the left and right column respectively, for different wall wavenumbers. For all panels, the tube position is included (as the top and bottom boundary curves) as indicated in figure to illustrate the relative position between the wall and thread interface. The panels in the first row show the case with a straight tube $\sigma=0$. For the next three panels, $k_w=3, 13, 23$ respectively. The initial Bloch wavenumber in the initial condition is fixed for all panels as $k=7$. Other parameters are $\epsilon=0.1, \sigma=0.2, \lambda_0=1, \bar{R}_e=0$.}
   \label{drop2}
\end{figure}

\begin{figure}
  \centerline{ \includegraphics[width=6.in,height=3.in]{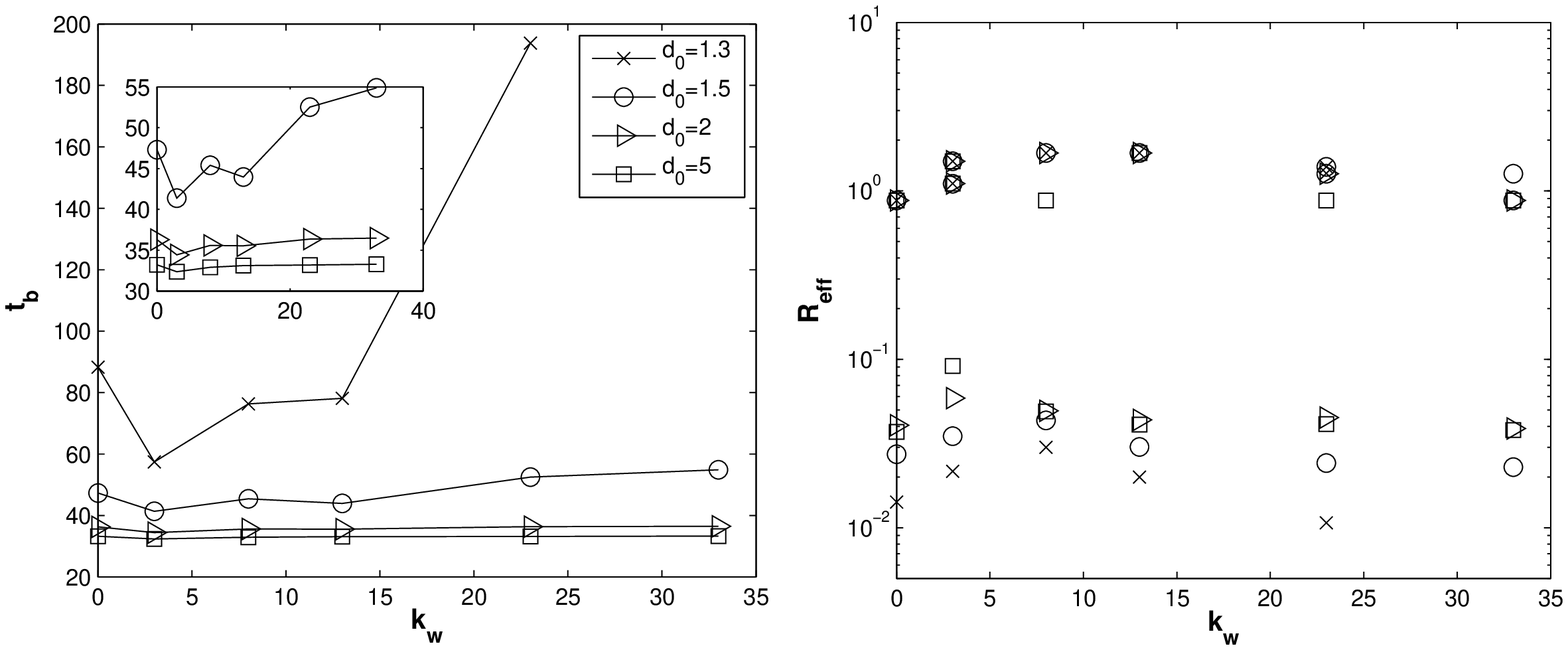} }
   \caption{The influence of wall wavenumber $k_w$ on the breakup times and drop sizes for the pinching solutions. The initial wavenumber $k=7$ and $\sigma=0.2$ are chosen as in figure \ref{drop1} and \ref{drop2}. The values at $k_w=0$ are those corresponding to the straight tube cases ($\sigma=0$) which are used to compare with the cases when the tube wall is corrugated. The inset in the left panel shows more details of the variation for $d_0=1.5,2,5$. The symbols that are used in the right panel are for the same cases indicated in the left panel.}
   \label{phs2}
\end{figure}

In this subsection, we investigate the drop formation toward the breakup of the liquid threads or layers. In figure \ref{drop1}, drop and satellite formations for various wall shapes are shown when the tube wall is initially far away from the thread interface, $d_0=5$, so that the tube wall is out of sight in all the panels (the same tube wall shapes will be shown in figure \ref{drop2}). In the top panels of figure \ref{drop1}, the tube wall is kept uncorrugated, $\sigma=0$, while $k_w=3,13,23$ with $\sigma=0.2$ for the following three ones from the top to the bottom. It is seen that the drop formation can be quite different when the wall has wavy structure. In the panels in the second row, $(k,k_w)=(7,3)$ corresponds to a point that leads to a larger linear growth rate than the straight tube case (see panel (a) in figure \ref{linear}), which causes the thread to pinch with fewer large or mother drops than the top panel. The satellite formations are illustrated in the right column of figure~\ref{drop1}.
In addition, further short-wave perturbation causes the surface of large drop to have dimples. When $k_w=13$, the panel (b) of figure \ref{linear} shows the linear growth rate dominated by the resonant one $|k-k_w|=6$ which is close to $k=7$, while $k_w=23$, one unstable branch completely covers the growth rate curve in the case of no wall corrugation. Therefore, it is expected that the eventual drop shapes in the third and bottom row are similar to the top panel. 
However, a close inspection reveals that, although the drop shape is similar, the detail of pinch-off differs. For $\sigma=0$, the satellites are obtained simultaneously since the actual period is $2\pi/7\approx 0.9$ in the top row of figure~\ref{drop1}. For $\sigma=0.2$ and $k_w=13$ in the third row, the middle satellite is obtained via the first breakup while the rest possible satellites can only appear from the subsequent breaks up (since the necking region has already formed, it is likely they will be the satellite drops; see the related work in \cite{Stone1992jfm} which presented a 'self-repeating' mechanism on a series of breakups to obtain satellites). In the bottom row, the left (right) most satellite together with the middle satellite breakups earlier than the ones next to them.
The calculation after the first pinch off may be interesting, but this is beyond the scope of this paper.
The difference in droplet sizes (together with the cases having different values of $d_0$) will be summarized later in figure~\ref{phs2}.

%However, a close inspection reveals that, although drop shape is similar, the position of the drops at breakup differs, or the length of the thin thread connecting large drops is different. For example, the pinching point in the middle thread is at $z\approx 0.18$ in the top panel, while $z\approx 0.2$ in the third panel and $z\approx 0.21$ in the bottom panel. In addition, the drop toward the end of period (at around $z\approx 2.2\sim 2.3$ for example) is shifted in space. 

Since the tube wall would appear far away from the local pinching region (assuming the tube cross section variation is not too large), the pinching behavior would be expected to resemble the case with no wall corrugation (see the study on the pinching dynamics in \cite{Wang2013}). Therefore this is not pursued further in current work.

\begin{figure}
  \centerline{ \includegraphics[width=6.in,height=3.in]{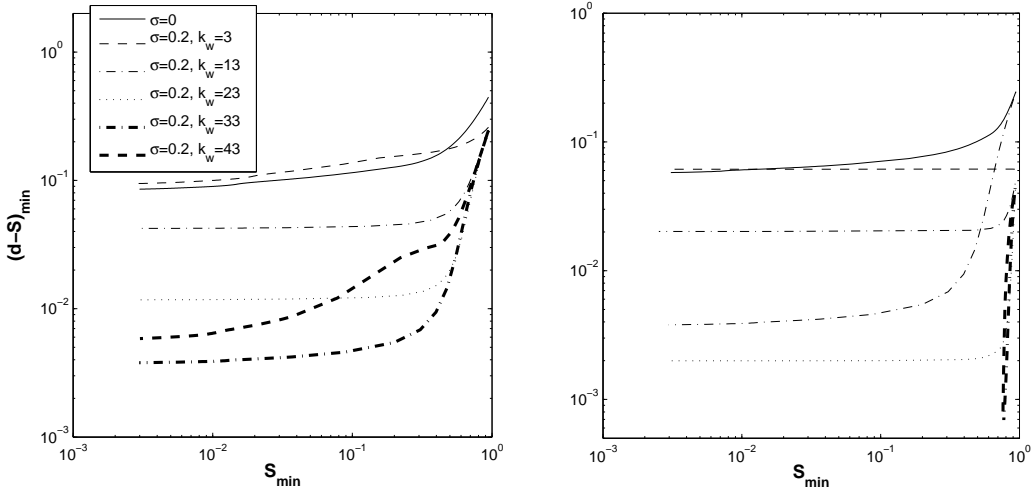} }
   \caption{The evolution of the minimum gap between the thread interface and the wall versus the minimum core thread neck for $d_0=1.5$ in the left panel and $d_0=1.3$ in the right panel. Here $(d-S)_{min}$ is defined as the difference in the radial direction because of the one dimensional model, rather than the local normal distance to the tube wall.
   In the right panel, it is obvious that for $k_w=33,43$, the gap becomes small while the neck of core thread $S_{min}$ remains $O(1)$.}
   \label{film0}
\end{figure}

\begin{figure}
  \centerline{ \includegraphics[width=6.in,height=4.in]{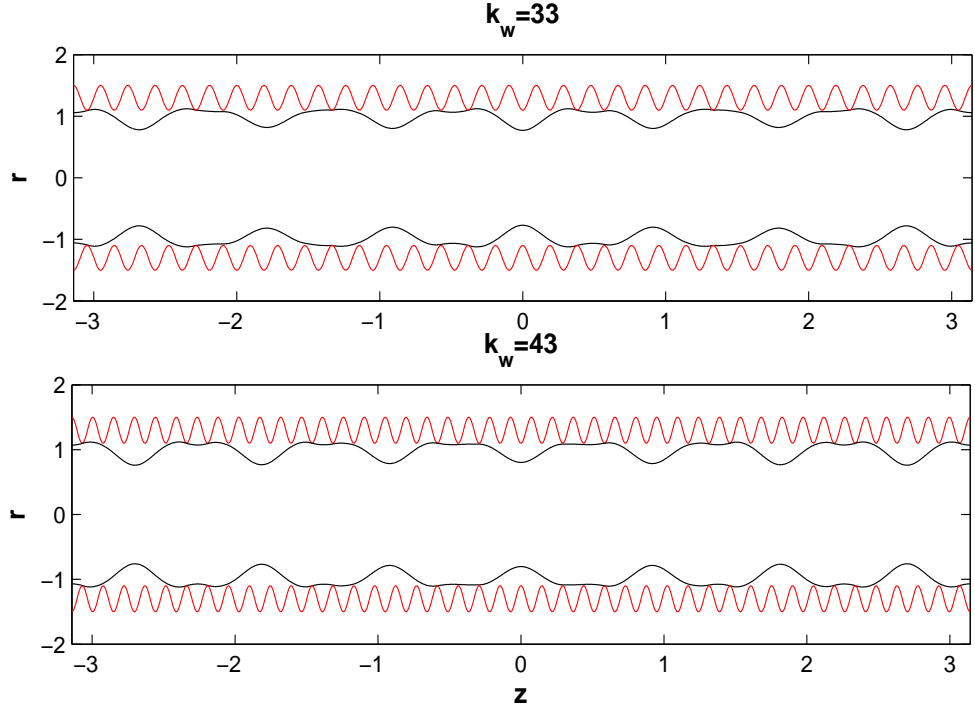} }
   \caption{The thread interface shapes for annular film drainage when $d_0=1.3, k_w=33, 43$. The top panel is for time $t=257.1$ while $t=417.3$ at the bottom panel.}
   \label{film1}
\end{figure}

When the annular layer is relatively thin, namely, the averaged tube radius $d_0$ is small, we show drop formation for $d_0=1.5, 1.3$ in figure \ref{drop2}, where the wall shapes are chosen to be the same as in figure \ref{drop1} but they are visible in figure now. In the top two panels, the tube is straight, $\sigma=0$ and we obtain the so called plug-with-collar drop formation (see \cite{Hagedorn2004} and \cite{Wang2013}) where the fluids are trapped between the tube wall and the large drops along with pinch off. When the tube wall has variation in shape, the thread needs to cope with the wall constraint and larger drops can be expected at first breakup even starting with the same initial perturbation wavenumber. The drop shapes resemble the results in \cite{Olgac2006} in the full axisymmetric simulations when the wall variation is present. The panels in the second row of figure \ref{drop2} show larger drops in the middle of the thread compared to the top panels where the tube wall is straight, because a longer wave length perturbation is excited by the wall harmonics, $|k-k_w|=4<k=7$ (see also the argument in \cite{WR2002a}). In the third row, pinching occurs near $z=0$, leaving a long thread both at $z>0$ and $z<0$. Additionally, the left one in the third row would perhaps experience a second breakup due to the thin neck around $z\approx \pm 1$, which leads to multiple drop formations.
Meanwhile, in some portion of the tube, the annular layer tends to breakup as well, since as pinching occurs, the film drainage regime is also approached slowly. Due to the wall topography, it is seen that more fluids compared to the case $\sigma=0$ can be trapped between the wall and drops.
In the last row of figure \ref{drop2}, drop formation is seen to differ significantly. In the left panel, a ultimate multiple drop formation is expected, while in the right panel, a large size drop is obtained in the middle which is connected by drops in comparable sizes. 
%Meanwhile the drop in the last panel from the right column of figure \ref{drop2} is shown to be very large and pinching point is away from $z=0$. 
As shown later, this case in the right bottom panel stands near a transition from pinching solutions to film rupture solutions.

The associated evolutions of the corresponding minimum thread radius and the minimum thread-tube gap are plotted in figure \ref{film0}, where the left panel for the case $d_0=1.5$, shows as $S_{min}\rightarrow 0$, i.e. the core thread pinches, the minimum gap still remain significantly larger compared to $S_{min}$ or at least, $S_{min}$ decays faster than the gap value. Therefore pinching solutions are obtained (at least up to $k_w=43$ here).
As shown in the thin annulus limit, owing to the different time scales (recall $\tau=\delta^3 t$ with $\delta\ll 1$ in (\ref{heqn}) and (\ref{xieqn})), pinching is expected to occur in advance of film drainage that happens in longer time dynamics, and perhaps even after the first pinch off, as shown in \cite{Hagedorn2004} in the straight uncorrugated tube case. In addition, we observe from the left panel of figure \ref{film0} that the change in minimum gap along pinching with $k_w$ is not monotonic. It seems that the gap remains larger for $k_w=43$ than $k_w=33$.
The right panel of figure \ref{film0} shows similar results to the left panel, except that $d_0=1.3$ and for sufficiently large $k_w$ ($k_w>23$), the neck of core thread $S_{min}$ remains order one while the gaps seem to reach a very small value first.
%For all the values of $d_0\ge 1.25$ that are investigated here, the pinching solution is always obtained first. 
% to be confirmed
In those cases, we provided the numerical evidence that a tube wall with large wavenumber may induce the film drainage regime, meanwhile suppress pinching even for a case where the annulus is not asymptotically thin. 
In the case without wall corrugation, $\sigma=0$, to suppress pinching, sufficiently thin annulus is required (the numerical work by \cite{Pozrikidis92jfm} gave a threshold value $d_0\approx 1.19$ for the uncorrugated tube in Stokes flow); in contrast, we found $d_0=1.3$ (namely, the average film thickness $0.3$) is sufficiently small to produce this phenomenon. In general, this should also depend on the size of wall variation $\sigma$ (as well as the initial conditions on the thread profiles). For the discussion here, $\sigma=0.2$ is fixed all the time and we merely provide the evidence of possible pinching suppression for an annular layer whose thickness is not asymptotically small.
Figure \ref{film1} shows the thread interface shape at the final stage of our simulations, at $t=257.1$ and $t=417.3$ respectively, where it seems that the effective tube radius is reduced ($\sim d_0-\sigma$) due to the rapid variation of tube cross section, which makes the fluids inside the wall undulation effectively rigid. Notice that in figure \ref{film1}, the core fluid does not intrude the inside of the wall undulation part, while in figure \ref{drop2} (the third and fourth row in particular), the core fluid still can wrap around the undulation 'tips'.
%In the case of $d_0=1.5$, pinching solutions seem always obtainable, even for $k_w=33,34$, for which the annulus has sufficient room to evolve to be close to the tube wall before pinching occurs. 
%However, these solution behaviors depend, in general, on the initial conditions. For example, one can choose a relatively large perturbation in the initial thread profile so that the interface is close enough to the tube wall
%This shows that for a tube wall with sufficiently large wavenumber, or with sufficient roughness, the pinching solution can be 
%but such computation is time consuming and is not pursued further here; figure \ref{linear_compr} shows the growth rate can be as small as $10^{-5}$ when $d_0=1.1$. 
%For all the values of $d_0\ge 1.16$ that are investigated here, the pinching solution is always obtained. The numerical work of the full axisymmetric problem in \cite{Pozrikidis92jfm} gives a threshold value $d_0\approx 1.19$ below which pinching can be suppressed in the case with a uncorrugated wall i.e. the dynamics is dominated by the Hammond equation. 
%However, we have not found such transition in solutions from pinching to film drainage. As shown in \ref{thinfilm}, the latter occurs in asymptotically small thin annulus.
%We do not pursue this further to establish this transition boundary in our long wave model.
As the liquid layer keeps thinning, locally, one still expects the lubrication model holds as well as the scaling laws based on it (see \cite{Hammond1983} and \cite{LRKCJ2006}, where $h\sim t^{-1}$ for a collar next to a collar, $h\sim t^{-1/2}$ for a lobe next to a collar, with $h$ the minimum film thickness to the substrate and $t$ the time). Unfortunately our numerical results fail to show the scalings even the gap is small and in the order of $O(10^{-4})$. It is possible that the thin film regime has not been fully reached yet as the fluid interface is only touching the wall undulation tip and the fluid region adjacent to the thinning part is different from the collar or lobe that is seen in \cite{LRKCJ2006}. 

Finally, for pinching solutions, we show the effect of the wall wavenumber $k_w$ on the breakup times and drop sizes, with the latter characterized by an effective drop radius $R_{eff}$ that is defined as
\begin{equation}
R_{eff} = \left(\frac{3}{4}\int_{z_1}^{z_2} S^2 dz\right)^{1/3},
\end{equation}
where $z_1, z_2$ are two points associated with sufficiently thin thread neck that we consider to be the pinching points.
In the left panel of figure \ref{phs2}, the breakup time $t_b$ first decreases and then increases as $k_w$ varies between $0$ and $10$. This is expected from our linear theory (also qualitatively similar to the results in figure \ref{phs1}). 
When $k-k_w$ corresponds to a mode with larger growth rate, the breakup time is smaller. For example, when $d_0=1.5$, the dominant mode is $k=4$ for $k_w=3$ and $k=7$ or $1$ for $k_w=8$. Accordingly, the growth rate is $0.06$ and $0.049$ respectively, which indicates the former case breakups earlier than the latter one.
%explains the difference in $t_b$. 
As $k_w$ increases, $t_b$ increases and the increasing is more profound when $d_0$ is smaller. For $d_0=2, 5$, the breakup time eventually approaches the value in the straight tube case for large $k_w$ approximately, because the initial wavenumber is already a long wave one, and one of the unstable branches in the linear theory for $\sigma>0$ has almost covered the straight tube case (see figure \ref{linear_compr}). 
While for the relatively small $d_0$, the annular film also tends to reach the drainage regime which occurs in a longer time scale and tends to suppress the pinching. As given in figure \ref{film1}, the annular film touching solution is obtained instead of pinching for $k_w=33$ and the interface profile that is shown is at $t=257.1$.
The right panel of figure \ref{phs2} illustrates the distribution of a large and satellite drops. 
For clarity here, we only calculate the satellite size from the first breakup (typically the one satellite in the middle) together with the adjacent large drops, where we estimated a second pinching point, as our current code can only track down to the first breakup. But for $k_w=3, d_0=5$ in figure \ref{drop1}, there are clearly three large drops and two small satellites, which are all taken into account. This has also been done for the small $d_0$ cases. 
For $d_0=5$ in figure \ref{drop1}, multiple satellite formation is expected eventually, which is similar for $d_0=2$ (data not shown). 
%For example for $k_w=13$ in figure \ref{drop1}, more satellites are expected after the first breakup since the slender neck(s) connecting large drops tend to form. We suspect the process will be similar to the results described in \cite{Stone1992jfm}, namely, every pinch-off is associated with the formation of some necking region which breakups subsequently and the process repeats. 
%But for clarity here, we only calculate the satellite size from the first breakup (typically the one satellite in the middle and we estimated a second pinching point) together with the adjacent large drops as our current code can only track down to the first breakup, except for $k_w=3$, where there are clearly three large drops and two small satellites, which are all taken into account. This has also been done for the small $d_0$ cases. 
As $k_w$ increases, it is seen that the size of both the large and satellite drops increases to a local maximum for some $k_w$ within the range $0<k_w<10$, or $0<\epsilon k_w<1$, which depends on the value of $d_0$. The drop sizes are similar to the straight tube case when $k_w$ is relatively large except when $d_0$ is sufficiently small. For $d_0=1.3$, the size of the satellite drop keeps decreasing as $k_w$ increases. Based on our numerical results in figure \ref{film0}, this indicates a transition from the pinching to touching solution.

% conclusion
\section{Concluding remarks}\label{sec:concl}
We have studied the effect of wall corrugation on the linear stability and nonlinear dynamics of viscous liquid threads and annular layers inside a cylindrical tube. A long wave model accounting for the wall shape effect, as well as the coupling between the annular layer and an active core thread, has been derived, which, in appropriate limits, reduces to the equations in literature \cite{EgDu1994,Hammond1983,WR2002a, WR2002b, ListerStone98}. 
%{\red Two main features of the current model are: the coupling between an active core and liquid layer, and taking into account of the wall topography. This distinguishes our study from the usual modeling works which have modeled either a liquid film by neglecting the core dynamics or a single liquid jet by neglecting the effect of external fluid.}
Similar to the findings in \cite{WR2002a}, the linear stability of the system shows that the short waves, which would be stable in the uncorrugated tube case, can excite unstable long waves through the interaction with the wall shape variations. When the wall wavenumber is relatively small, along with unstable short waves, the long waves are more unstable, or have larger dominant growth rate, compared to the uncorrugated wall case. Consequently, the breakup time is expected to decrease.
A stable band of modes appear when the wall wavenumber is sufficiently large. 

The stability of the threads and layers in the nonlinear regime was then investigated via direct numerical simulations of the evolution equations. Starting with relatively small initial perturbation, the results in general agree with our linear theory with a transition period, where the growth first follows the value predicted in the uncorrugated tube problem. Drop formation as well as drop size is shown to alter by including the wall corrugation. Typically, larger and fewer drops can be expected based on the simulation results, by choosing proper parameters $d_0$ and $k_w$. Meanwhile, similar to the 'plug with collar' formation (see \cite{Hagedorn2004,Wang2013}), accompanied by pinching, fluids are found to be trapped between the tube wall and the core thread, where the film drainage regime is approached and is expected to persist perhaps even after the first pinch off. Furthermore, we showed possibility that pinching solutions can be suppressed depending on the choices of $d_0$ and $k_w$.

The model here provides a template for further studies or extensions; one can include some external field (gravity, electric field e.g.) or interfacial surfactants, as much remains to be explored. Meanwhile, this reduced model has the advantage over the full problem regarding the computational cost issue.
Furthermore, with this model, other more complicated wall shapes can be incorporated easily, which are related to the engineering design for heat transfer optimization, drug or particle delivery and emulsification (see \cite{Utada2005} e.g.) in microfluidic devices and also in biological systems \cite{JBG2011,LHFG2014}. We already have a few preliminary results underworking on controlling the droplet sizes by changing channel or tube wall shapes. Intuitively for example, liquid thread breaks slower in a tube with smaller radius because of a smaller growth rate. Therefore a combination or a contraction-expansion tube would lead to a series of droplets in the expansion side, when the thread is not broken yet in the contraction or narrow side, which is similar to those 'flow-focusing-device' ideas to some extent, although the roughness of the wall could potentially complicates the dynamics. Similarly, with locally dimpled tube, different size of droplets can be obtained so that a desired drop distribution may be achieved. Flow rate is another key parameter in the experiments and this will be investigated in the future by inserting axial flow or some external forcing in our model. However, our present work focuses only on the stability of a stationary thread as the axial flow does not affect the real part of our growth rate $\omega$, of temporal stability. We take this as a first step of modeling to understand the physics of the system and more are under working.
\\

\Appendix
\section{Formulation of the linearized problem}
In this section, we briefly derive the formulation for calculating the linear growth rate by applying the FFH method.
Let $S=e^{\omega t}\tilde{S}=e^{\omega t+ik z}\psi(z)$ and $g=(4\lambda_0G)^{-1}$, where $k$ is the Bloch wavenumber or Floquet multiplier; then (\ref{llw}) becomes
\begin{equation}
\omega\left(\tilde{S} -6g_z\tilde{S}_z-6g\tilde{S}_{zz}\right) = - \left(g\left(\tilde{S}_z+\epsilon^2\tilde{S}_{zzz}\right)\right)_z\label{eig1}
\end{equation}
Multiplying the above equation (\ref{eig1}) by $e^{-ik z}$ and taking the Fourier transform, we obtain the following equation (more details can be found in \cite{Deconinck2006})
\begin{align}
\omega\mathcal{L}^l(k)\hat{\psi} = \mathcal{L}^r(k)\hat{\psi} \label{geigen}
\end{align}
with $\hat{\psi}= (\cdots, \hat{\psi}_{-2}, \hat{\psi}_{-1}, \hat{\psi}_0, \hat{\psi}_1,\hat{\psi}_2,\cdots)^T$ and
where 
\begin{equation}
\psi(z) = \sum_{j=-\infty}^{\infty}\hat{\psi}_j e^{ik_wjz},\quad
%\end{equation}
{\rm with}\quad
%\begin{equation}
\hat{\psi}_j=\frac{1}{2L}\int_{-L}^L \psi(z)e^{-ik_wjz}dz,\quad j\in \mathcal{Z}.
\end{equation}
In addition, the bi-infinite matrices are defined by
\begin{align}
\mathcal{L}_{nm}^i(k)=\sum_{l=0}^{M_i}\hat{f}^i_{j,n-m}\left[k+k_w m\right]^j 
\end{align}
where $i=l$ or $r$, $M_i$ is the highest number of derivatives in $i$th side and $f^i_{j}$ stands for the coefficient function in front of $j$th derivatives and can be represented by a Fourier series
\begin{equation}
f^i_{j}(z) = \sum_{l=-\infty}^{\infty}\hat{f}^i_{j,l} e^{ik_wlz},\quad
%\end{equation}
{\rm with}\quad
%\begin{equation}
\hat{f}^i_{j,l}= \frac{1}{2L}\int_{-L}^L f^i_{j}(z)e^{-ik_wlz}dz.
\end{equation}
In particular, $M_l=2$ and $M_r=4$ for (\ref{eig1}), and
\begin{align}
&f^l_0=1,\quad f^l_1=-6g_z,\quad f^l_2=-6g,\\
& f^r_0=0,\quad f^r_1= -g_z,\quad f^r_2= -g,\quad f^r_3=-\epsilon^2g_z\quad f^r_4=-\epsilon^2 g.
\end{align}
The spectrum is $k_w$ periodic and only $-\frac{k_w}{2}\le k\le \frac{k_w}{2}$ is needed to discuss since larger range gives no more information (see \cite{WR2002a} e.g.). To calculate the spectrum numerically, we choose a cut-off $N$ on the number of Fourier modes of the eigenfunctions $\psi$, resulting in a linear system of dimension $2N+1$ from (\ref{geigen}).
%\section{}

\bibliographystyle{siam}
\bibliography{Capjets}

\end{document}